\documentclass[fleqn,usenatbib]{mnras}
\usepackage{newtxtext,newtxmath,lscape}
\usepackage[T1]{fontenc}
\usepackage{lineno}
\usepackage{ulem}
\usepackage{multirow}

\DeclareRobustCommand{\VAN}[3]{#2}
\let\VANthebibliography\thebibliography
\def\thebibliography{\DeclareRobustCommand{\VAN}[3]{##3}\VANthebibliography}

\newcommand{\oiii}{[O\,{\sc iii}]} 

\newcommand{\civ}{C\,{\sc iv}} 
\newcommand{\heii}{He\,{\sc ii}}
\newcommand{\mgii}{Mg\,{\sc ii}}
 
\newcommand{\bhm}{M_{\bullet}}

\usepackage{graphicx}	
\usepackage{amsmath}	

\title[Searching for periodic quasar candidates from ZTF]{\bf Searching for quasar candidates with periodic variations from the Zwicky Transient Facility: results and implications}

\author[Chen et al.]
{Yong-Jie Chen$^{1,2}$, 
Shuo Zhai$^{1,3}$,
Jun-Rong Liu$^{1,3}$,
Wei-Jian Guo$^{4}$,
Yue-Chang Peng$^{1,3}$,
Yan-Rong Li$^{1}$\thanks{E-mail: liyanrong@mail.ihep.ac.cn},	
\newauthor
Yu-Yang Songsheng$^{1,2}$,
Pu Du$^{1}$, 	
Chen Hu,$^{1}$,
Jian-Min Wang$^{1,4,5}$\thanks{E-mail: wangjm@mail.ihep.ac.cn}
	\\
	$^{1}$Key Laboratory for Particle Astrophysics, Institute of High 
	Energy Physics, Chinese Academy of Sciences, 19B Yuquan Road, Beijing 100049, China\\
        $^{2}$Dongguan Neutron Science Center, 1 Zhongziyuan Road, Dongguan 523808, People,s Republic of China\\
	$^{3}$School of Physical Science, University of Chinese Academy of Sciences, 19A Yuquan Road, Beijing 100049, People’s Republic of China \\
	$^{4}$National Astronomical Observatories of China, Chinese Academy of Sciences, A20 Datun Road, Beijing 100012, China\\
	$^{5}$School of Astronomy and Space Science, University of Chinese Academy of Sciences, 
	19A Yuquan Road, Beijing 100049, China
}

\date{Accepted XXX. Received YYY; in original form ZZZ}

\pubyear{2022}

\begin{document}
	\label{firstpage}
	\pagerange{\pageref{firstpage}--\pageref{lastpage}}
	\maketitle
	
	\begin{abstract}
        We conduct a systematic search for quasars with periodic variations from the archival photometric data of the Zwicky Transient Facility by cross-matching with the quasar catalogs of the Sloan Digital Sky Survey and V{\'e}ron-Cetty \& V{\'e}ron. We first select out 184 primitive periodic candidates using the generalized Lomb-Scargle periodogram and auto-correlation function and then estimate their statistical significance of periodicity based on two red-noise models, i.e., damped random walk (DRW) and single power-law (SPL) models. As such, we finally identify 106 (DRW) and 86 (SPL) candidates with the most significant periodic variations out of 143,700 quasars. We further compare DRW and SPL models using Bayes factors, which indicate a relative preference of the SPL model for our primitive sample. We thus adopt the candidates identified with SPL as the final sample and summarize its basic properties. We extend the light curves of the selected candidates by supplying other archival survey data to verify their periodicity. However, only three candidates (with 6-8 cycles of periods) meet the selection criteria. This result clearly implies that, instead of being strictly periodic, the variability must be quasi-periodic or caused by stochastic red-noise. This exerts a challenge to the existing search approaches and calls for developing new effective methods.
	\end{abstract}
	
	
	
	\section{Introduction}
	Periodic variability of quasars has attracted great attention over the past decade because of its possible connection with supermassive black hole binaries on the theoretic side (e.g., \citealt{Artymowicz1996,DOrazio2015}) and significant advances towards modern time-domain surveys on the observational side. There had been more than one hundred periodic quasar candidates reported from systematic searches over large surveys, such as from Catalina Real-time Transient Survey (CRTS, \citealt{Graham2015b}), Palomar Transient Factory (PTF, \citealt{Charisi2016}), Pan-STARRS1 Medium Deep Survey (PS1 MDS, \citealt{Liu2019}), and from a combination of the Dark Energy Survey (DES) and Sloan Digital Sky Survey (SDSS; \citealt{Chen2020}). Several individual quasars were serendipitously found to exhibit (quasi-)periodic variations using long-term archival data (e.g., \citealt{Valtonen2008,LiYR2016, LiYR2019, Zhang2020, ONeill2022, Zhang2021}). Those candidates constitute
	a precious guiding sample for investigating the physical origins of the periodicity and exclusively identifying supermassive black hole binaries. However, we bear in mind that the periodicity might be subject to false positives caused by red-noise stochastic variability of quasars (e.g., \citealt{Vaughan2016}). A sophisticated estimation of the statistical significance of the periodicity in the presence of red-noise variations is highly required. Meanwhile, a systematic search over new emerging time-domain surveys with improved sampling rates and photometry would provide more valuable candidate samples to testify to the periodicity. 

    The Zwicky Transient Facility (ZTF) is a new time-domain survey dedicated to a systematic exploration of the northern optical transient sky
    with a 47 square degree field of view camera equipped on the 48-inch Samuel Oschin Telescope at Palomar Observatory (\citealt{Bellm2019,Graham2019}). It started operation in 2018 and scans the entire northern visible sky with a typical cadence of three days. Such a moderate sampling cadence makes ZTF well-suitable for exploring quasar periodicity. In this work, we conduct a systematic search for periodic quasar candidates based on ZTF photometric data by cross-matching with the quasar catalogues of the SDSS and \cite{Veron2010}. 

    The paper is organized as follows. In Section~\ref{sec_sample_method}, we describe the search methodology and estimation of the statistical significance of the periodicity. In Section~\ref{sec_results}, we briefly summarize the basic properties of our selected periodic quasar candidates and compare our sample with those reported in previous works. In Section~\ref{sec_discussion}, we compile other available survey photometric data to extend the temporal baselines to verify the periodicity, and then present a brief discussion on possible explanations for the periodicity and implications of our periodic search. Finally, we summarize the main results in Section~\ref{sec_summary}.
    Throughout this work, we use a cosmology with $\Omega_{\rm M}=0.3$, $\Omega_\Lambda=0.7$, and $H_0=70\ \rm km\ s^{-1}Mpc^{-1}$.

	\section{Quasars Sample and Methods} 
	\label{sec_sample_method}
	\subsection{Initial Quasar Catalogues and ZTF Light Curve Data}\label{sec2.1}
	We initiate with the quasar catalogues of SDSS DR14 and \cite{Veron2010}. The former contains 526,356 spectroscopically identified quasars (\citealt{Paris2018})
	and the latter contains 168,941 quasars compiled from all known quasars at that time. To discard faint objects, we respectively set a magnitude limit of $r\leq20$ for SDSS DR14 catalogue and $V\leq20$ for \cite{Veron2010} catalogue. These limits are slightly brighter than the ZTF depth of $\sim20.5$ mag (\citealt{Graham2019}). We then perform a cross-match within a radius of 5 arcsecs between the two catalogues and remove duplicated objects using the package {\tt esutil}\footnote{\url{https://github.com/esheldon/esutil}.}. We combine the left objects in the two catalogues and obtain a list of 223,061 quasars.
	
	Next, we retrieve the above initial quasar list from the ZTF database using a searching radius of 3 arcsecs. We use the 11th ZTF public data release\footnote{The details of the 11th ZTF data release can be found at \url{https://irsa.ipac.caltech.edu/data/ZTF/docs/releases/dr11}.}. There are three custom filters, $g$-, $r$- and $i$-band. The photometry was reduced with the Infrared Processing and Analysis Centre pipeline (\citealt{Masci2019}). The $r$-band generally has the most data points, therefore, unless stated otherwise, we use the $r$-band light curves for the following analysis.
	
	In some light curves there apparently appear a few outlier points. We use a fifteenth-order polynomial to fit the light curves and remove those points outside the 3$\sigma$ deviation from the best fits. The polynomial order is adopted somewhat arbitrarily, but generally, it should be high enough to capture all major variation patterns in the light curves. Since we only focus on long-timescale variability, we bin the light curves within an interval of 20 days. The magnitudes and errors of the binned points are assigned the mean and standard error of the points in each 20-day bin, respectively. This binning operation helps to alleviate
	the possible biases arising from severely uneven sampling and strong short-timescale fluctuations in some light curves.
	We further discard those objects with less than 40 binned points and are finally left with 143,700 quasars. Below we use the binned light curves for preliminary periodicity searches in Section~\ref{sec_method}.

	\begin{table*}
            \setlength{\tabcolsep}{3pt}
            \renewcommand{\arraystretch}{1.2}
		\centering
		\caption{Properties of the initial 184 periodic quasar candidates identified in this work. The full version is available in a machine-readable form online.}
		\label{table1}
		\begin{tabular}{lccccccccccccc} 
        \hline
        ID & RA & Dec & $z$ & Magnitude & $P_{\rm GLS}$ & $P_{\rm ACF}$ & $N_{\rm cycle}$ & $\rm \xi$& $\log L_{5100}$ & $\log M_{\bullet}$ & $\log\dot{\mathscr{M}}$\\
          &  &   &  & $({\rm ZTF}-r)$ & (day) & (day) &  & &  & $({\rm egs\ s^{-1}})$ & $(M_{\odot})$ &  \\
        \hline
        J000458.79-022629.5 & 00:04:58.79 & -02:26:29.54 &  0.432 & 18.4 & $644.1\pm53.6$ & $665.8\pm0.9$ & 2.0 & 7.6 & $44.50\pm0.07$ & $8.34\pm0.23$ & $-0.44\pm0.48$\\
        CRSS0008.4+2034 & 00:08:26.50 & +20:34:32.16 &  0.389 & 17.0 & $605.3\pm71.5$ & $637.5\pm2.0$ & 2.1 & 8.6 & --- & --- & ---\\
        J002209.95+001629.3 & 00:22:09.95 & +00:16:29.31 &  0.575 & 18.2 & $650.1\pm45.3$ & $623.8\pm0.6$ & 1.9 & 4.9 & $44.96\pm0.05$ & $7.67\pm0.23$ & $1.59\pm0.47$\\
        J002500.42-031238.5 & 00:25:00.42 & -03:12:38.54 &  0.543 & 17.0 & $788.8\pm29.6$ & $850.6\pm1.9$ & 1.6 & 5.3 & $45.28\pm0.03$ & $8.53\pm0.24$ & $0.34\pm0.48$\\
        J002504.53+213224.6 & 00:25:04.53 & +21:32:24.63 &  1.669 & 18.2 & $736.2\pm38.6$ & $779.6\pm0.8$ & 1.7 & 4.8 & $46.35\pm0.02$ & $9.31\pm0.30$ & $0.14\pm0.60$\\
        J002815.26+201420.8 & 00:28:15.26 & +20:14:20.89 &  0.406 & 19.8 & $688.6\pm53.8$ & $650.1\pm0.7$ & 1.8 & 5.0 & $44.21\pm0.06$ & $8.20\pm0.23$ & $-0.59\pm0.47$\\
        J003456.70+242649.8 & 00:34:56.70 & +24:26:49.81 &  0.907 & 18.2 & $859.0\pm31.9$ & $826.8\pm1.5$ & 1.6 & 4.3 & $45.63\pm0.02$ & $8.66\pm0.30$ & $0.37\pm0.60$\\
        J003718.86+143221.9 & 00:37:18.86 & +14:32:21.95 &  0.798 & 18.5 & $839.4\pm44.8$ & $876.7\pm3.6$ & 1.5 & 7.2 & $45.41\pm0.03$ & $9.21\pm0.30$ & $-1.06\pm0.60$\\
        J003858.53+020132.8 & 00:38:58.53 & +02:01:32.81 &  1.043 & 18.7 & $736.2\pm24.7$ & $772.5\pm2.7$ & 1.7 & 5.3 & $45.29\pm0.06$ & $9.29\pm0.30$ & $-1.40\pm0.61$\\
        J004039.79+150321.2 & 00:40:39.79 & +15:03:21.21 &  0.884 & 18.2 & $626.6\pm51.1$ & $672.3\pm0.7$ & 2.0 & 6.1 & $45.36\pm0.03$ & $9.21\pm0.30$ & $-1.14\pm0.60$\\
        J004744.03+190338.5 & 00:47:44.03 & +19:03:38.55 &  2.010 & 18.4 & $688.6\pm44.6$ & $678.5\pm1.3$ & 1.8 & 4.9 & $46.40\pm0.04$ & $9.73\pm0.30$ & $-0.62\pm0.61$\\
        \hline
		\end{tabular}
	\end{table*}

        %
	\begin{figure*}
		\centering
		\includegraphics[width=0.85\textwidth]{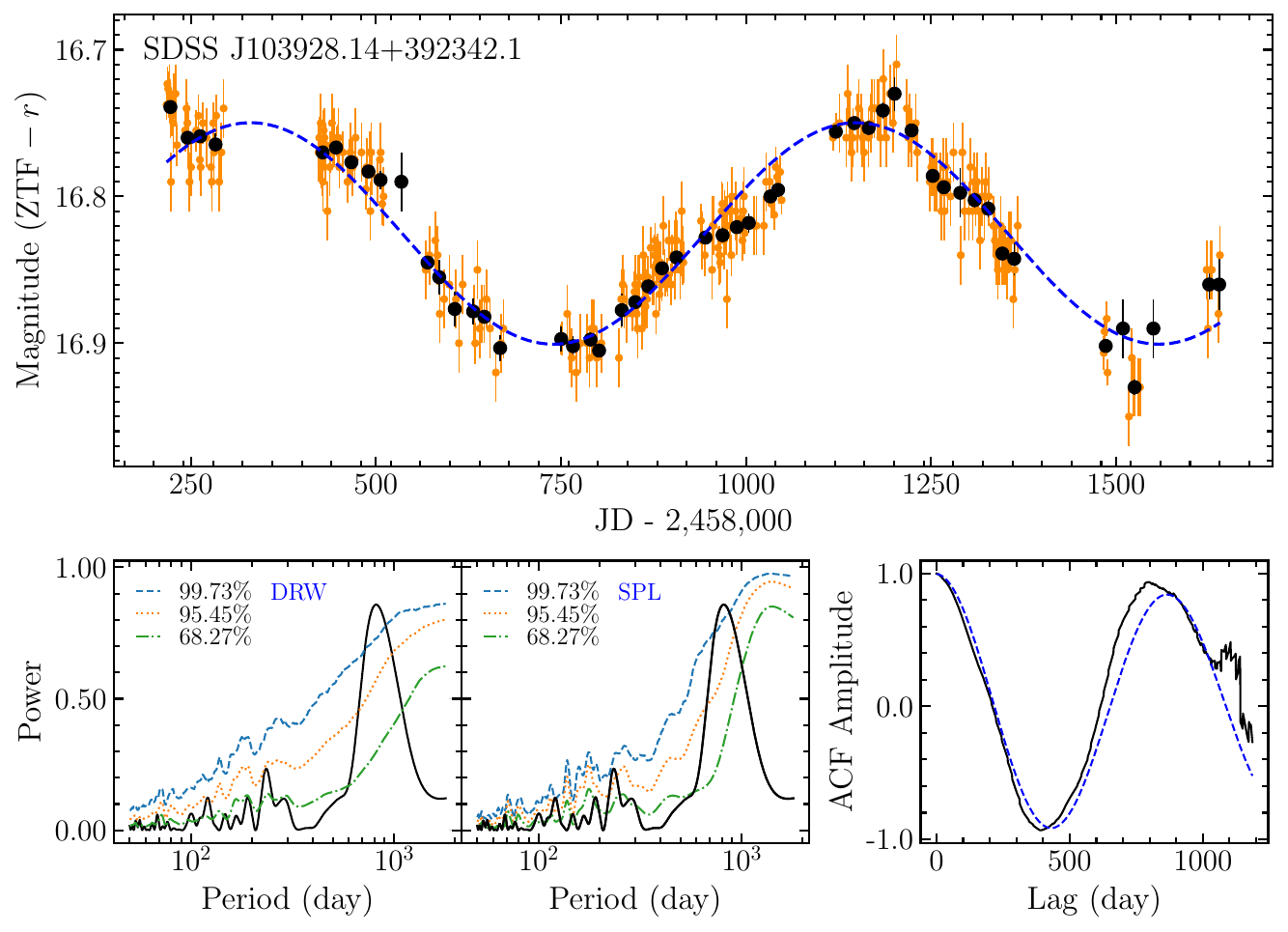}
		\includegraphics[width=0.85\textwidth]{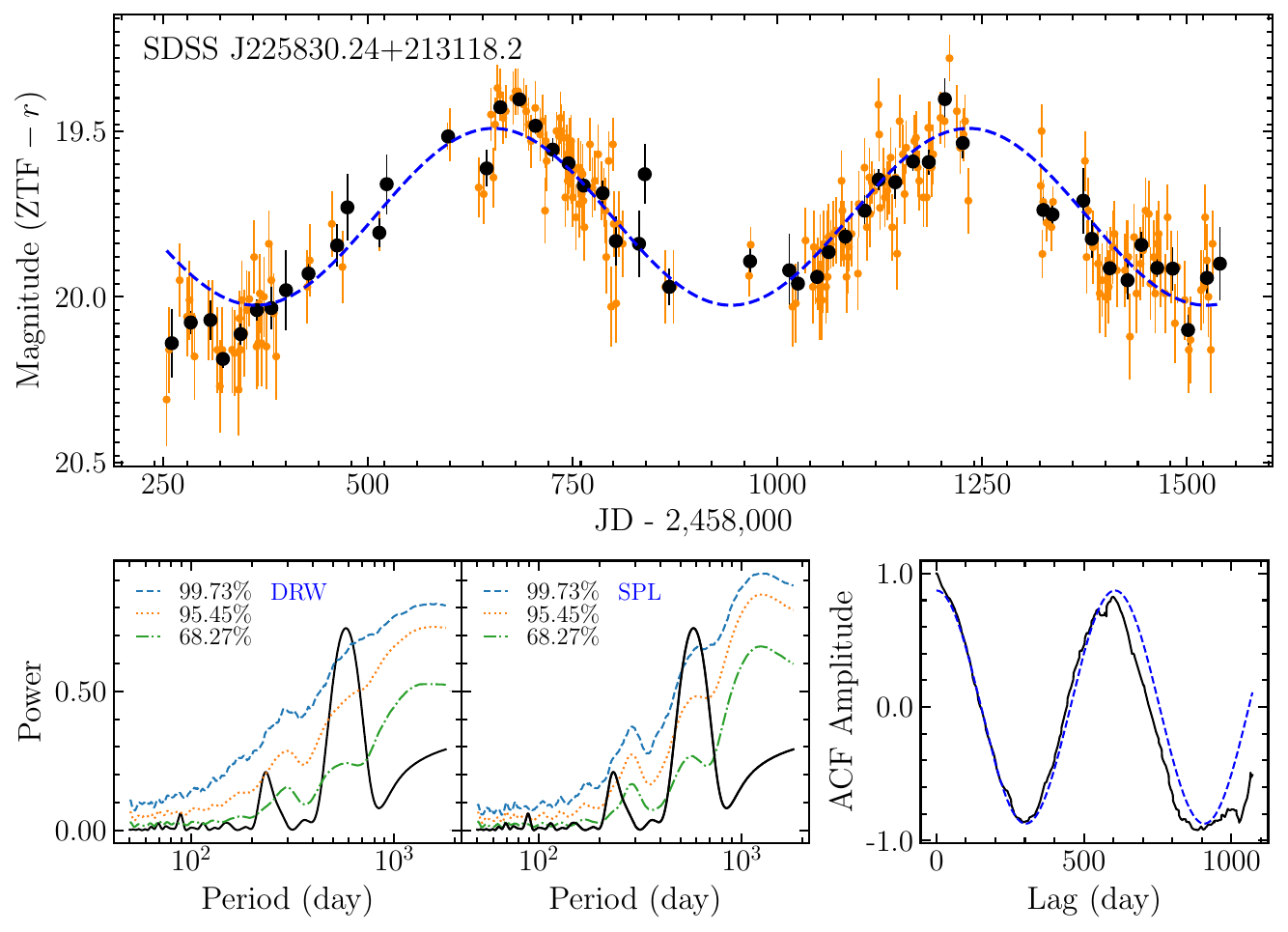}
		\caption{Two examples of our selected periodic quasar candidates. For each candidate, the upper panel shows the light curves from ZTF DR11. The orange points show the original $r$-band light curve and the black points show the binned light curve with an interval of 20 days. The blue dash line shows the best sinusoidal fit. The two bottom left panels show the GLSP and the significance levels in terms of DRW and SPL red-noise models (see Section~\ref{sec_significance_freq}). The bottom right panel shows the ACF of the light curve (the black solid line) and the best fit of an exponentially decaying cosine function (the blue dashed line; see Section~\ref{sec_method}). }
		\label{fig_lc}
	\end{figure*}

	\subsection{Identification Methodology of Periodicity}\label{sec_method}
	In this section, we design the following two steps to identify raw periodic quasar candidates. In the next section, we will select out the final candidate sample according to the estimated significance of the periodicity.

	First, we employ the generalized Lomb-Scargle periodogram (GLSP; \citealt{Zechmeister2009}) implemented in the {\tt pyastronomy}\footnote{\url{https://pyastronomy.readthedocs.io/en/latest/index.html}.} package.
	The Lomb-Scargle periodogram (LSP) was proposed by \cite{Lomb1976} and \cite{Scargle1982} and had become a widely used method to detect periodic signals in unevenly sampled data. It is mathematically equivalent to a sinusoidal fit with a form of 
	\begin{equation}
	M(t) = a\sin\left(\frac{2\pi t}{P}+\phi\right),
	\label{eqn_sin}
	\end{equation}
	where  $a$ is the amplitude, $P$ is the period, and $\phi$ is the phase. Compared to the LSP, the GLSP additionally includes an offset term in the sinusoidal fit. The advantage of this generalization is that it provides a more accurate period prediction and a better determination of the spectral intensity \citep{Zechmeister2009}.

	We calculate the GLSP over a period range between 50 days and $T_{\rm span}$, where $T_{\rm span}$ is the time span of the light curve.
	We assign the best period $P_{\rm GLS}$ corresponding to the largest periodogram peak and then perform a sinusoidal fit with the period fixed to $P_{\rm GLS}$.
	As such, we obtain the amplitude $a_0$ of the best-fit sinusoidal function and the residuals by subtracting the best-fit sinusoidal function from the light curve.
	Following \cite{Horne1986}, we define a signal-to-noise (S/N) ratio as $\xi=a_0^2/2\sigma_r^2$, where $\sigma_r$ is the standard deviation of the residuals.
	This ratio measures the power arising from the periodic signal relative to the power from the noise. We require the periodic candidates to satisfy the criterion: 1) $\xi>4.0$ so as to select out the statistically significant peak; and 2) $N_{\rm cycle} = T_{\rm span}/P_{\rm GLS}>1.5$ to ensure at least 1.5 cycles of periodicity in the light curve.

	Second, we adopt the auto-correlation function (ACF) to search for periodicity by combining it with the GLSP.
	ACF describes the degree of auto-correlation of the light curve at different time lags and has also been widely used for detecting periodic signals (e.g., \citealt{McQuillan2013, Graham2015b, Chen2020}).
	Theoretically, the ACF of periodically driven stochastic systems is expected to follow an exponentially decaying cosine function \citep{Jung1993}, i.e., 
	\begin{equation}
	{\rm ACF(\tau)} = a\cos(2\pi\tau/P_{\rm ACF})\exp(-\lambda\tau),  
	\label{eqn_acf}
	\end{equation} 
	where $\tau$ is the time lag and $P_{\rm ACF}$ is the period. We use the method of the interpolated cross-correlation function (ICCF; \citealt{Gaskell1987,White1994}) to calculate the ACF and fit it with the exponentially decaying cosine function. We require that 1) the period difference between the GLSP and ACF methods is within 10 percent, namely, $\left|1-P_{\rm ACF}/P_{\rm GLS}\right|\textless 0.1$; and 2) the best-fit decay rate $\lambda \textless 10^{-3}$ day$^{-1}$, which ensures the ACF decaying no more than $1/e$ over the temporal baseline (about 1500 days) of the light curve (\citealt{Graham2015b}).

	We obtain 184 periodic candidates that satisfy the above selection criterion (see Tabel~\ref{table1} for details). In Fig.~\ref{fig_lc}, for the sake of illustration, we show the original and binned light curves, along with the GLSP and ACF of the binned light curve for two selected examples. We note that Vaughan et al. (2016) suggested searching over light curves with at least three-period cycles to distinguish between true candidates for periodicity versus light curves that can be fully described via standard stochastic red-noise processes. For these 184 candidates, the present temporal baselines of ZTF data are not long enough to ensure this criterion (the range of period cycles is from 1.5 to 2.8). In Section 4.2, we will supplement these baselines with earlier archival data from other surveys to extend the temporal baselines for selected candidates.

    \begin{figure*}
        \centering
        \includegraphics[width=0.85\textwidth]{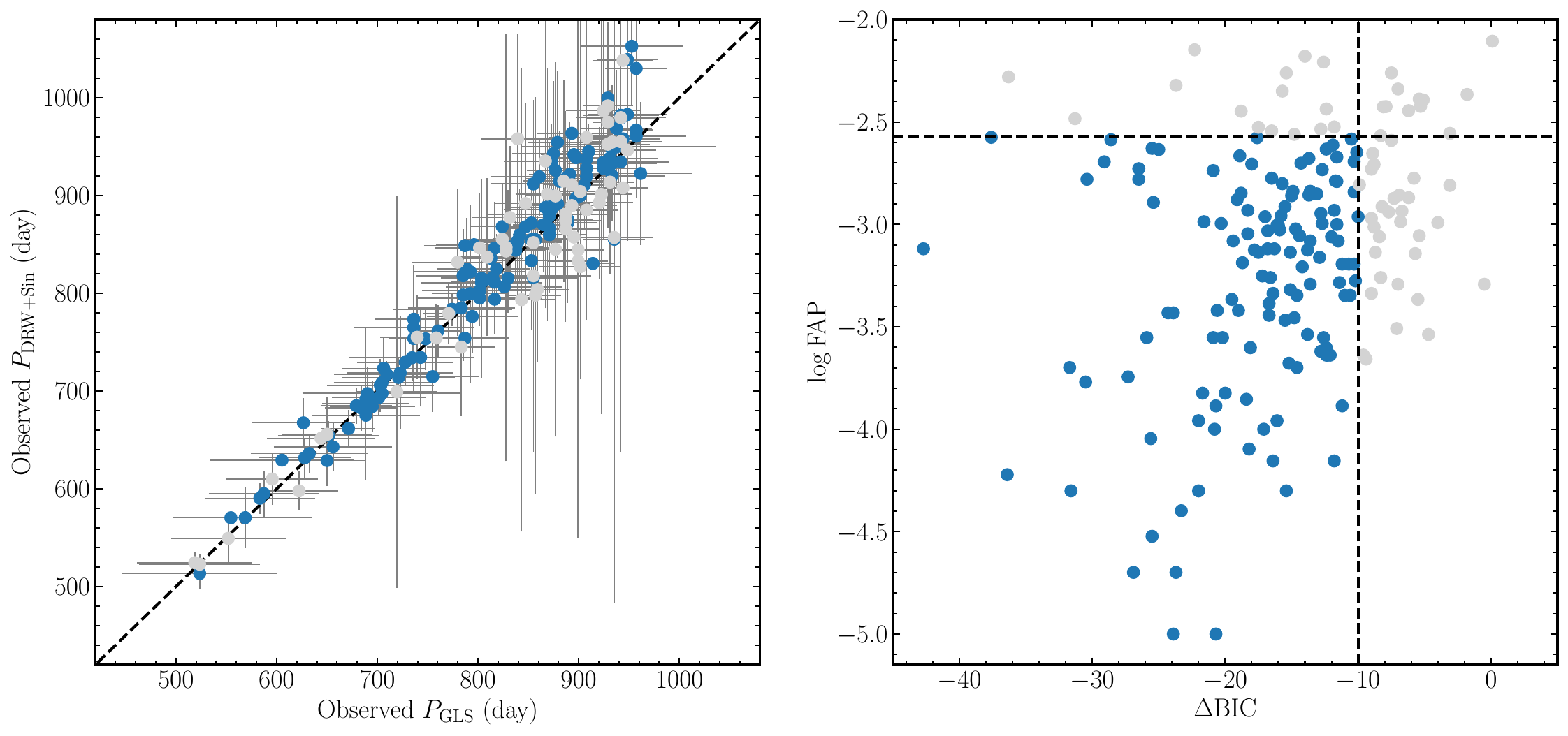}
        \caption{Left panel: a comparison between the periods from the DRW+sinusoidal model fitting and the periods from the GLSP. Right panel: a comparison between the FAP and $\rm \Delta BIC$. The horizontal and vertical dotted lines correspond to (1-FAP) = 99.73\% and $\Delta \rm BIC = -10$, respectively. In each panel, the blue points highlight those candidates with $\rm (1-FAP)\geq99.73\%$ and $\Delta \rm BIC \leq -10$.}
        \label{fig_modelselection}
    \end{figure*}

	\subsection{Estimating Significance of the Periodicity}\label{sec_significance}
        It is known that red-noise-like variability is commonly seen in quasars/active galactic nuclei (AGNs)\footnote{We use the terminology ``quasar'' and ``AGN'' interchangeably throughout the paper.} can produce spurious few-cycle periodic signals in AGN light curves, especially for those with sparse and uneven sampling and large photometric uncertainties (e.g., \citealt{Vaughan2016,Krishnan2021}). Considering that we only limit 1.5 cycles of periodicity, it is necessary to appropriately estimate the statistical significance of the periodicity. 
        Below we estimate the false-alarm probability (FAP) and significance based on two popular red-noise models, namely, the damped random walk (DRW) model and the single power-law (SPL) model. Here, the SPL model means the power spectral density (PSD) of the light curve follows a SPL. 
    
	\subsubsection{The False-alarm Probability with the DRW Model}\label{sec_fap}
		
	We adopt the DRW process \citep{Kelly2009} as the null hypothesis. The DRW process is widely used to describe AGN stochastic variability (e.g., \citealt{Kelly2009,Zu2011,Li2013,Li2016,Lu2019,Zhang2022}) and has a covariance function in the form of
    \begin{equation}
    S_{ij} = \sigma_d^2\exp\left(-\frac{\left|t_i-t_j\right|}{\tau_d}\right), 
    \label{eqn_sij}
    \end{equation} 
    where $t_i$ and $t_j$ are two times, $\sigma_d$ is the variation amplitude at long timescale, and $\tau_d$ is the characteristic damping timescale. For each periodic candidate, we fit the original unbinned light curve by exploring the posterior probability function 
    \begin{equation}
    \ln P(X|\sigma, \tau, q) \propto \ln \mathcal{L} + \ln P(\sigma, \tau, q),
    \end{equation}
    where $q$ is the long-term mean of the light curve, $P(\sigma, \tau, q)$ is the prior probability function, and $\mathcal{L}$ is the likelihood function, defined as 
    \begin{equation}
    \mathcal{L} \varpropto |C|^{-\frac{1}{2}}\exp\left[-\frac{1}{2}\sum_{i, j}(X_i-q)(C^{-1})_{ij}(X_j-q)\right],
    \label{eqn_likeli}
    \end{equation}
    where $X_i$ represents the observed light curve. The covariance matrix $C$ is given by
    \begin{equation}
	    C_{ij} = \sigma_{i}^2\delta_{ij} + S_{ij},
	\end{equation}
	where $\sigma_i$ is the measurement uncertainty at the observation time $t_i$, $\delta_{ij}$ is the Kronecker delta, and $S_{ij}$ is given by Equation~(\ref{eqn_sij}). 
    
    We set a uniform prior for the logarithm of the parameters $\sigma_d$ and $\tau_d$ and a uniform prior for $q$. We apply the Markov Chain Monte Carlo (MCMC) sampler \texttt{emcee}\footnote{\url{https://emcee.readthedocs.io/en/stable/}.} \citep{Foreman2013} to construct the posterior samples of the parameters. 
    We then randomly draw a set of $\sigma_d$ and $\tau_d$ from their posterior samples and generate a simulated light curve with exactly the same cadence as the observed data. The parameter $q$ is a nuisance for the present purpose as we shift the generated mock light curve to match the observed mean magnitude. We further add Gaussian noises with a zero mean and standard deviation equal to the photometric uncertainties to mimic the measurement errors. Again, we bin the mock light curve every 20 days and apply exactly the same identification methods as in Section~\ref{sec_method}.

    We repeat the above procedures 100,000 times and calculate the FAP\footnote{This FAP has taken into account the \textit{look-elsewhere} effect~\citep{Algeri2016} arising from searches over a broad period range by counting for all possible false positives within the whole period range.} as ${\rm FAP}=N_{\rm p}/N_{\rm tot}$, where $N_{\rm p}$ is the number of periodic candidates in the mock light curves and $N_{\rm tot}$ is the total number of simulated light curves. 
    The FAP of the 184 candidates ranges from $1.0\times10^{-5}$ to $7.8\times10^{-3}$ (see the right panel of Fig.~\ref{fig_modelselection}).
    We mark a significant periodicity if $\rm (1-FAP)\geq99.73\%$, namely, at least a significance of 3$\sigma$.
	
    \subsubsection{The Bayesian Information Criterion of Periodicity with the DRW Model}
    \label{sec_bic}
    We have estimated the FAP with the null hypothesis that the light curves follow the DRW process. As an alternative approach, we perform a model selection to test if the data favor an additional periodic signal on top of the background DRW variability. 
	To this end, we adopt the Bayesian information criterion (BIC) defined as (\citealt{Schwarz1987})
	\begin{equation}
	    {\rm BIC} = -2\ln\mathcal{L}+k\ln N,
	\end{equation}
	where $\mathcal{L}$ is the likelihood function, $k$ is the number of free model parameters, and $N$ is the number of data points. As in Section~\ref{sec_method}, we simply adopt a sinusoidal model for the periodic signal and compare such a DRW+sinusoidal model with the pure DRW model. 
	
	The pure DRW model has three parameters, namely, the red-noise amplitude ($\sigma_d$), characteristic timescale ($\tau_d$), and mean magnitude ($q$). The DRW+sinusoidal model has three additional parameters, namely, the period ($P$), phase ($\phi$), and amplitude ($a$) of the periodic signal (see Equation~\ref{eqn_sin}). Based on Equation~(\ref{eqn_likeli}), the likelihood function for the DRW+sinusoidal model is written as 
	\begin{equation}
	\mathcal{L} \varpropto |C|^{-\frac{1}{2}}\exp\left[-\frac{1}{2}\sum_{i, j}(X_i-M_i-q)(C^{-1})_{ij}(X_j-M_j-q)\right],
	\end{equation}
	where $M_i$ represents the sinusoidal signal.
	We again employ the MCMC sampling package \texttt{emcee} to obtain posterior samples of the parameters, from which the best-estimated values and uncertainties are assigned by the means and standard deviations, respectively.
	
	In the left panel of Fig.~\ref{fig_modelselection}, we compare the best estimated periods with those determined by the GLSP in Section~\ref{sec_method}. We can find the general consistency within uncertainties, despite a few discrepant points that might be caused by the inclusion of the DRW process. The period uncertainties of both approaches generally increase with the period because a longer period means fewer cycles of periodic variations, resulting in fewer constraints on the period.
	We calculate the BIC difference between the above two models as 
	\begin{equation}
	\Delta{\rm BIC} = {\rm BIC_{DRW+Periodic}} - {\rm BIC_{DRW}}.    
	\end{equation} A negative $\Delta{\rm BIC}$ indicates a more preference over the DRW + sinusoidal model. As shown in the right panel of Fig.~\ref{fig_modelselection}, the obtained $\Delta{\rm BIC}$ are overall negative, meaning that the DRW+sinusoidal model is relatively more preferable. The right panel of Fig.~\ref{fig_modelselection} also plots the relation between $\Delta{\rm BIC}$ and the FAP estimated in the preceding section. There appears a generic positive correlation as expected if regardless of the scattering.  {It is noteworthy that the sinusoidal function exhibits a simple and well-defined structure, whereas a sinusoidal-shape light curve stemming from red-noise may exhibit slight deviations from the ideal sinusoidal form. Consequently, even if the fake periodicity is caused by red-noises, there is a possibility of favoring the DRW + sinusoidal model in our fitting. Therefore, it is crucial to carefully choose an appropriate BIC value in order to select the preferable models. To be conservative,}
	in addition to the criterion (1-FAP) $\geq 99.73\%$, we also require $\Delta{\rm BIC}\leq-10$\footnote{A value of $\Delta{\rm BIC}\leq-10$ indicates very strong evidence in favor of $\rm BIC_{DRW+Periodic}$ \citep{Raftery1995}.} to select out the best periodic candidates. In the end, we censor 57 less significant quasars from the 184 candidates identified in Section~\ref{sec_method} and obtain a sample of 127 periodic quasar candidates.

    \begin{table*}
        \setlength{\tabcolsep}{4pt}
        \renewcommand{\arraystretch}{1.2}
        \centering
        \caption{Statistical significance estimation of the periodicity. The full version is available in a machine-readable form online.}
	\label{table_siginificance}
	\begin{tabular}{lccccccccccccc} 
        \hline
        & \multicolumn{5}{c}{DRW} & &\multicolumn{3}{c}{SPL}\\
        \cline{2-6}
        \cline{8-10}
        ID & $\log(\tau_d/\rm day)$ & $\log(\sigma_d/\rm mag)$ & $\log\rm FAP$ & $\Delta$BIC & >3$\sigma$ in GLSP$^a$ & & $\beta$ & $\log\rm FAP$ & >3$\sigma$ in GLSP$^a$ & $\log B$  & HD$^b$ & HS$^b$\\
        \hline
        J000458.79-022629.5 & $2.72_{-0.25}^{+0.25}$ & $-0.65_{-0.12}^{+0.12}$ & -2.53 & -17.5 & Y & &3.01 & -2.10 & N & 2.72 & N & N\\
        CRSS0008.4+2034 & $2.70_{-0.24}^{+0.25}$ & $-1.07_{-0.12}^{+0.12}$ & -3.12 & -42.7 & Y & &3.23 & -2.49 & Y & 5.83 & Y & N\\
        J002209.95+001629.3 & $2.60_{-0.26}^{+0.30}$ & $-1.01_{-0.12}^{+0.14}$ & -2.35 & -15.7 & Y & &3.00 & -2.44 & Y & 2.25 & N & N\\
        J002500.42-031238.5 & $2.67_{-0.26}^{+0.27}$ & $-1.27_{-0.12}^{+0.13}$ & -2.84 & -13.6 & Y & &2.55 & -2.96 & Y & 0.38 & Y & Y\\
        J002504.53+213224.6 & $2.62_{-0.28}^{+0.29}$ & $-1.36_{-0.12}^{+0.14}$ & -3.00 & -16.0 & Y & &2.98 & -3.59 & Y & 1.04 & Y & Y\\
        J002815.26+201420.8 & $2.39_{-0.27}^{+0.37}$ & $-0.73_{-0.12}^{+0.17}$ & -2.84 & -10.3 & N & &2.31 & -2.41 & N & -0.79 & N & N\\
        J003456.70+242649.8 & $2.67_{-0.26}^{+0.28}$ & $-1.05_{-0.12}^{+0.13}$ & -2.42 & -7.9 & Y & &2.69 & -2.44 & Y & -0.08 & N & N\\
        J003718.86+143221.9 & $2.64_{-0.25}^{+0.27}$ & $-1.02_{-0.11}^{+0.14}$ & -2.95 & -12.8 & Y & &3.01 & -2.44 & Y & 0.89 & Y & N\\
        J003858.53+020132.8 & $2.71_{-0.27}^{+0.25}$ & $-0.97_{-0.12}^{+0.12}$ & -2.61 & -11.9 & Y & &3.08 & -3.13 & Y & 1.47 & Y & Y\\
        J004039.79+150321.2 & $2.67_{-0.26}^{+0.27}$ & $-1.08_{-0.12}^{+0.13}$ & -2.74 & -20.9 & Y & &3.19 & -2.67 & Y & 2.45 & Y & Y\\
        J004744.03+190338.5 & $2.48_{-0.28}^{+0.35}$ & $-1.30_{-0.12}^{+0.16}$ & -3.12 & -13.8 & N & &2.71 & -3.39 & Y & 0.23 & N & Y\\
        \hline
	\end{tabular}
        \begin{list}{}{}
        \item[$^a$]{Whether the highest peaks of GLSPs reach 3$\sigma$ significance levels. Y-yes and N-No.}
        \item[$^b$]{Whether the candidates satisfy our whole selection criteria for high significance when using the DRW (HD) and SPL (HS) models as the null hypothesis. Y-yes and N-No.}
        \end{list}
	\end{table*}

    \begin{figure}
    \centering
    \includegraphics[width=0.49\textwidth]{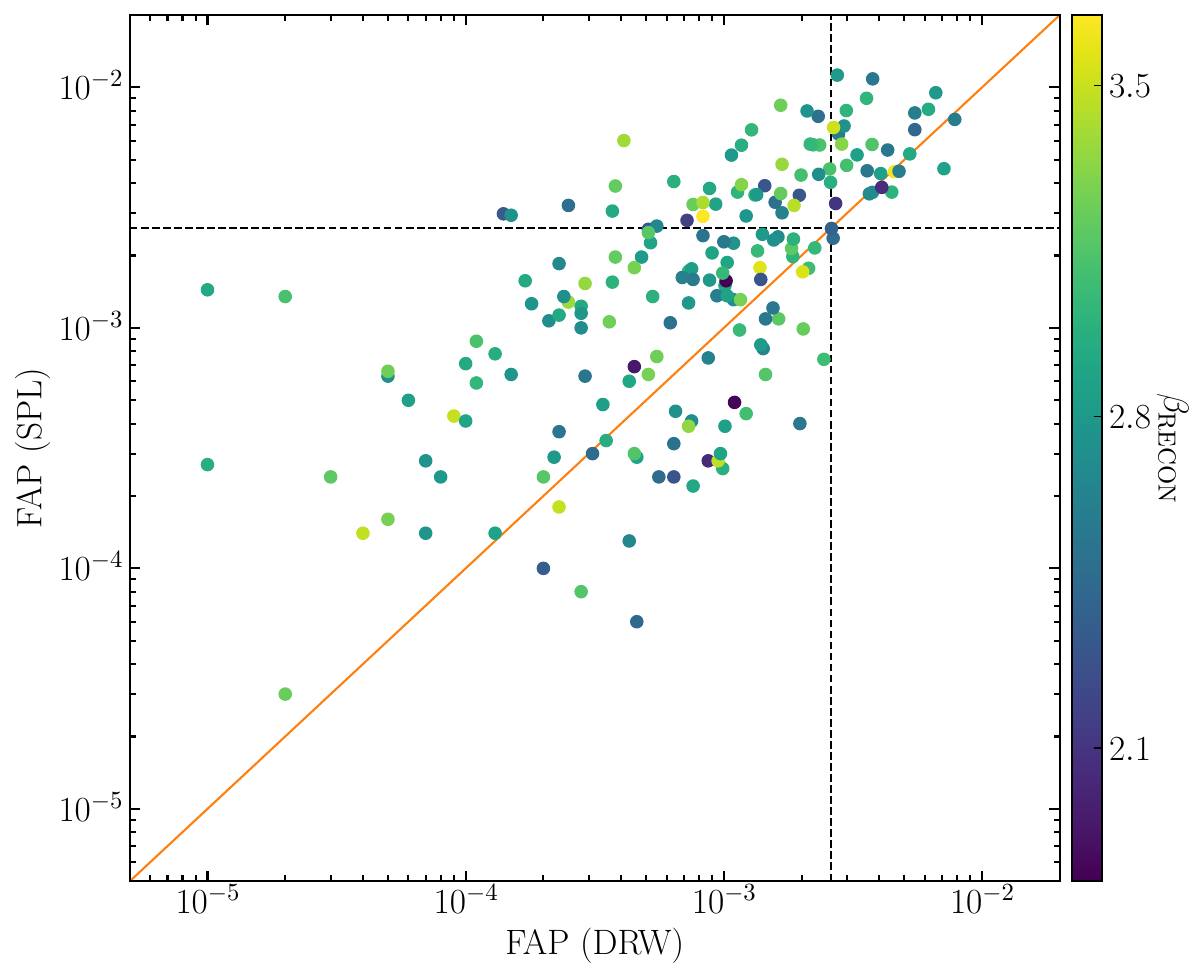}
    \caption{A comparison of FAPs between the DRW and SPL models. The points are color-coded by the best-fit slopes of the SPL from {\tt RECON}. Both the horizontal and vertical dashed lines correspond to (1-FAP) = 99.73\%.}
    \label{fig_FAP_Comparison}
    \end{figure}

    \subsubsection{The False-alarm Probability with the SPL Model}
    \label{sec_fap_pl}
    {In the preceding two sections, we assume that the AGN stochastic variability is characterized by the DRW process. 
    In this section, we estimate FAPs of the 184 candidates selected in Section~\ref{sec_method} using the SPL model. 

    Due to irregular sampling and seasonal gaps, the calculated PSD from the Fourier transform of the light curves may be distorted from the true spectra (\citealt{Uttley2002}; see a simple test in Appendix~\ref{sec_psd_test}), making it difficult to obtain reliable model parameters by directly fitting the PSDs. To overcome this limitation, we utilize the framework {\tt RECON}\footnote{\url{https://github.com/LiyrAstroph/RECON}.} (\citealt{Li2018}; see Appendix ~\ref{sec_psd_test} for a brief description of {\tt RECON} and validity tests.) to determine the parameters of the SPL model. {\tt RECON} uses a stochastic complex series to parameterize AGN variability in the frequency domain and transform the series back to the time domain to fit the observed light curve. {\tt RECON} can cope with irregularly sampled light curves and also can recover any forms of PSDs. The model parameters are determined using the MCMC technique so that their posterior samples can be thereby obtained. Here, the prior probabilities for the amplitudes\footnote{Prior to RECON analysis,  light curves are normalized by dividing their respective mean fluxes, therefore, the amplitudes here are indeed dimensionless.} and slopes of SPL are independently assigned as natural logarithmic and uniformly distributed, covering a range of (-15, 6) and (1, 5) respectively.
    
    After determining SPL parameters with {\tt RECON},
    we calculate the FAP using the same procedure as in Section~\ref{sec_fap}. We generate mock light curves with the method described by \cite{Timmer1995}. Fig.~\ref{fig_FAP_Comparison} compares FAPs derived from the DRW and SPL models. The obtained SPL slopes (listed in Table~\ref{table_siginificance}) are generally steeper than the DRW slope ($=$2) at high frequency (short timescale $<\tau$). Overall, the SPL model tends to yield relatively higher FAPs than the DRW model. This is not surprising as the SPL model results in higher red-noise powers at low frequencies (long timescale), which can easily produce more spurious periodic signals. Using the FAPs based on the SPL model, the number of AGNs that satisfy (1-FAP) $>$ 99.73\% decreases from 158 (DRW) to 120 (SPL).

    \begin{table}
        \renewcommand{\arraystretch}{1.2}
        \setlength{\tabcolsep}{21pt}
        \caption{The selection criterion used to identify periodic quasars candidates with high significance and the respective counts derived using the DRW and SPL models as the null hypothesis. The $\Delta$BIC is calculated only for the DRW model.\label{table_selection}}
        \begin{tabular}{lccc}
        \hline
        Selection Criterion & DRW & SPL  \\
        \hline
        Initial periodic candidates &  184 & 184\\
        (1-FAP)$>99.73\%$ & 158 & 120 \\
        $\Delta \rm BIC<-10$ & 141 & ---\\
        $>99.73\%$ in GLSPs & 144 & 124 \\
        Combined & 106 & 86\\
        \hline
        \end{tabular}
    \end{table} 
    
    }

    \subsubsection{Estimation of Significance with the GLS Periodogram }
    \label{sec_significance_freq}

    We also adopt the procedure proposed by \cite{Vaughan2005,Vaughan2010} to quantify the significance of the periodicity, which applies in the frequency domain and involves calculating the PSD of the light curve. However, as mentioned above, due to irregular sampling and seasonal gaps of the light curves, the PSDs obtained through direct Fourier transform are usually seriously distorted (see details in Appendix~\ref{sec_psd_test}). Therefore, we slightly modify the procedure outlined in \cite{Vaughan2005,Vaughan2010}. Specifically, instead of using PSDs, we employ the GLSP that applies to irregular sampling (see also \citealt{Chen2020}). By generating a series of mock light curves from DRW and SPL models, we can construct the respective background levels of red-noises in the GLSP. For each candidate, model parameters are randomly drawn from the corresponding posterior samples obtained by using the \texttt{RECON} framework. We then compare the peak power in the GLSP of the observed light curve with the background red-noise power levels to constrain the significance of the periodicity.

    In Fig.~\ref{fig_lc}, we plot 1$\sigma$ (68.27\%), 2$\sigma$ (95.45\%), and 3$\sigma$ (99.73\%) significance levels of DRW and SPL red-noises for two exemplary candidates. 
    For the DRW model as the null hypothesis, we identify 145 candidates with the peak powers in the GLSPs reaching the 3$\sigma$ significance level. By combining with the selection criterion outlined in Section~\ref{sec_fap} and \ref{sec_bic}, we obtain 106 high-significance candidates. For the SPL model, we find 86 periodic quasar candidates that have peak powers in the GLSPs reaching the 3$\sigma$ significance level (out of 138 objects) and (1-FAP) $>$ 99.73\%. We summarize these numbers of candidates with different selection criteria in Table~\ref{table_selection}.

    \section{RESULTS}
    \label{sec_results}
    \subsection{Comparison between DRW and SPL models}
    \label{sec_competition}
     The {\tt RECON} framework can calculate the Bayesian evidence and therefore allows us to compare different red-noise models based on the Bayes factor.
     For this purpose, we also run {\tt RECON} with the DRW model on the light curves of the selected candidates. We confirm that the obtained DRW parameters are consistent within uncertainties with those obtained through maximizing the likelihood in Equation~\ref{eqn_likeli} in Section~\ref{sec_fap}. We define the Bayes factor $B$ as the ratio of the Bayesian evidence calculated using the SPL and DRW models. A value of  $B>1$ means the SPL model is preferable to the DRW model. A widely used criterion for quantifying a strong preference in Bayesian model selection provided by \cite{Kass1995} is $\log B>1.3$.
     
\begin{figure}
    \centering
    \includegraphics[width=0.42\textwidth]{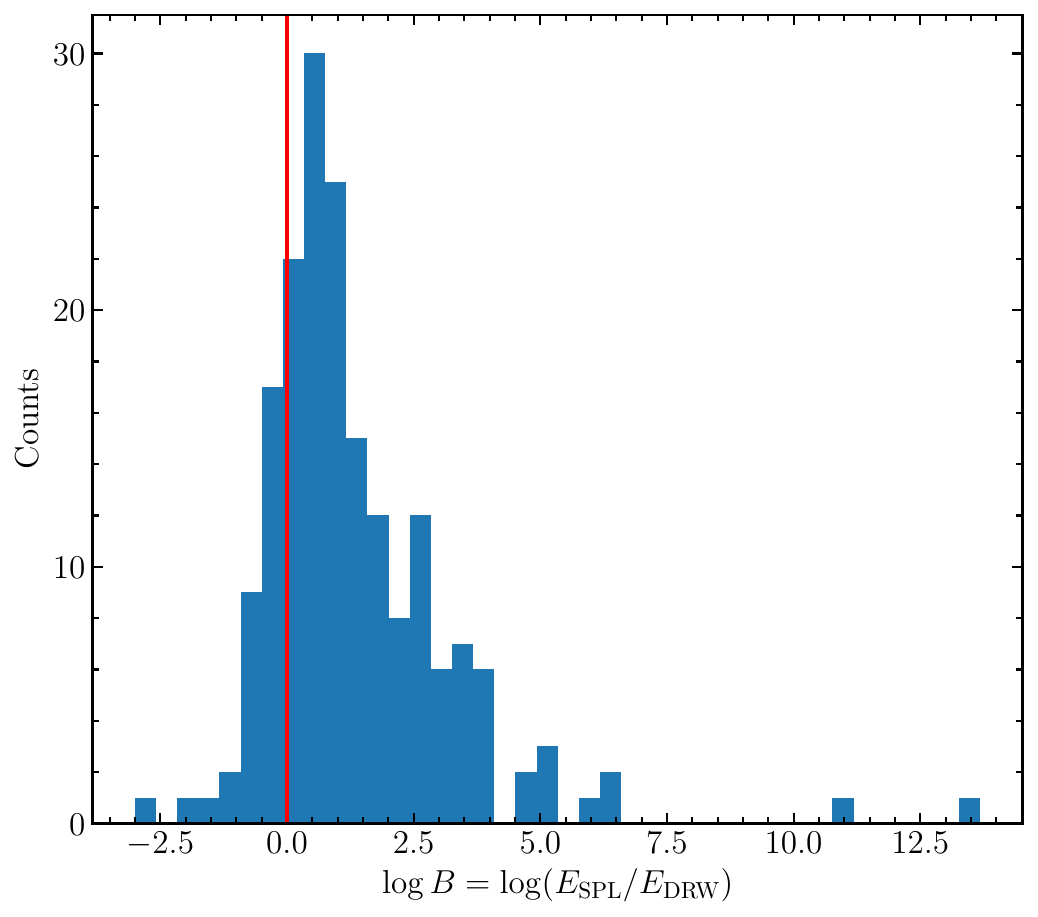}
    \caption{The distribution of Bayes factors $\log B$, defined as the evidence ratio of the SPL to DRW models (i.e., $B = E_{\rm SPL}/E_{\rm DRW}$, where $E_{\rm SPL}$ and $E_{\rm DRW}$ are the Bayesian evidence). The vertical red solid line shows $\log B=0$.}
    \label{fig_distribution_bayesfactor}
    \end{figure}

     Fig.~\ref{fig_distribution_bayesfactor} shows the distribution of $\log B$ for the 184 periodic candidates with a median value of $\log B\approx0.94_{-1.1}^{+1.8}$. This is a marginal value, indicating that the two models exhibit comparable goodness-of-fit to the observed data.
     Considering that most of the candidates (151/184) have $\log B>0$ (i.e., the SPL model is relatively more favorable) and the SPL model gives a more conservative significance estimation\footnote{It is worth noting that the best-fit parameter $\tau_{d}$ of DRW may be underestimated due to the limited time spans of ZTF light curves, leading to an overestimation of significance of the periodicity.}, we prefer the sample of 86 periodic candidates with high statistical significance identified using the SPL model. Below we by default use this sample as the fiducial periodic sample and proceed with a summary of its properties and further discussions.

    \begin{figure*}
        \centering
        \includegraphics[width=0.95\textwidth]{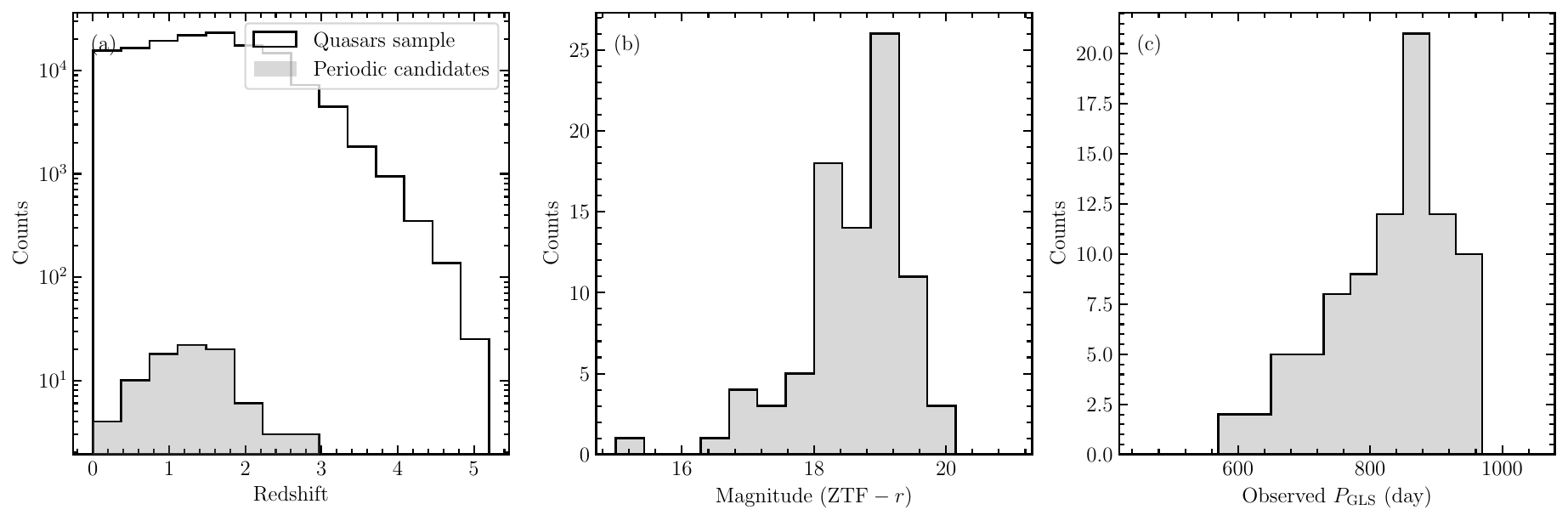}
        \caption{The distributions of the redshift, magnitude, and observed period (from the GLSP) of the periodic quasar sample. Note that the vertical axis of the rightmost panel is in a logarithm scale.}
        \label{fig_hist}
    \end{figure*}

    \begin{figure*}
    \centering
    \includegraphics[width=0.95\textwidth]{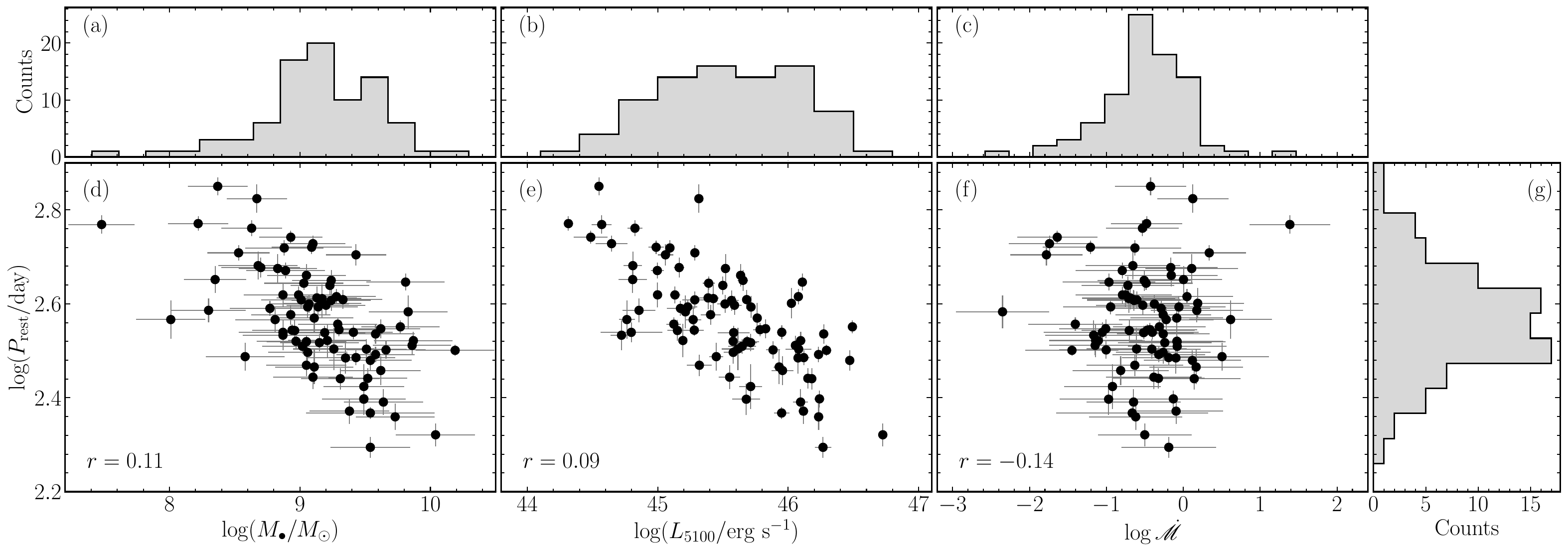}
    \caption{The rest-frame period (from the GLSPs) versus black hole mass, 5100~{\AA} luminosity, and accretion rate of our periodic quasar sample. Note that the superficial anti-correlations are caused by the redshift effect. The denotation $r$ shows the partial correlation coefficient excluding the dependence on redshift (see the text). } 
    \label{fig_properties}
    \end{figure*}

    \begin{figure*}
	\centering
	\includegraphics[width=0.9\textwidth]{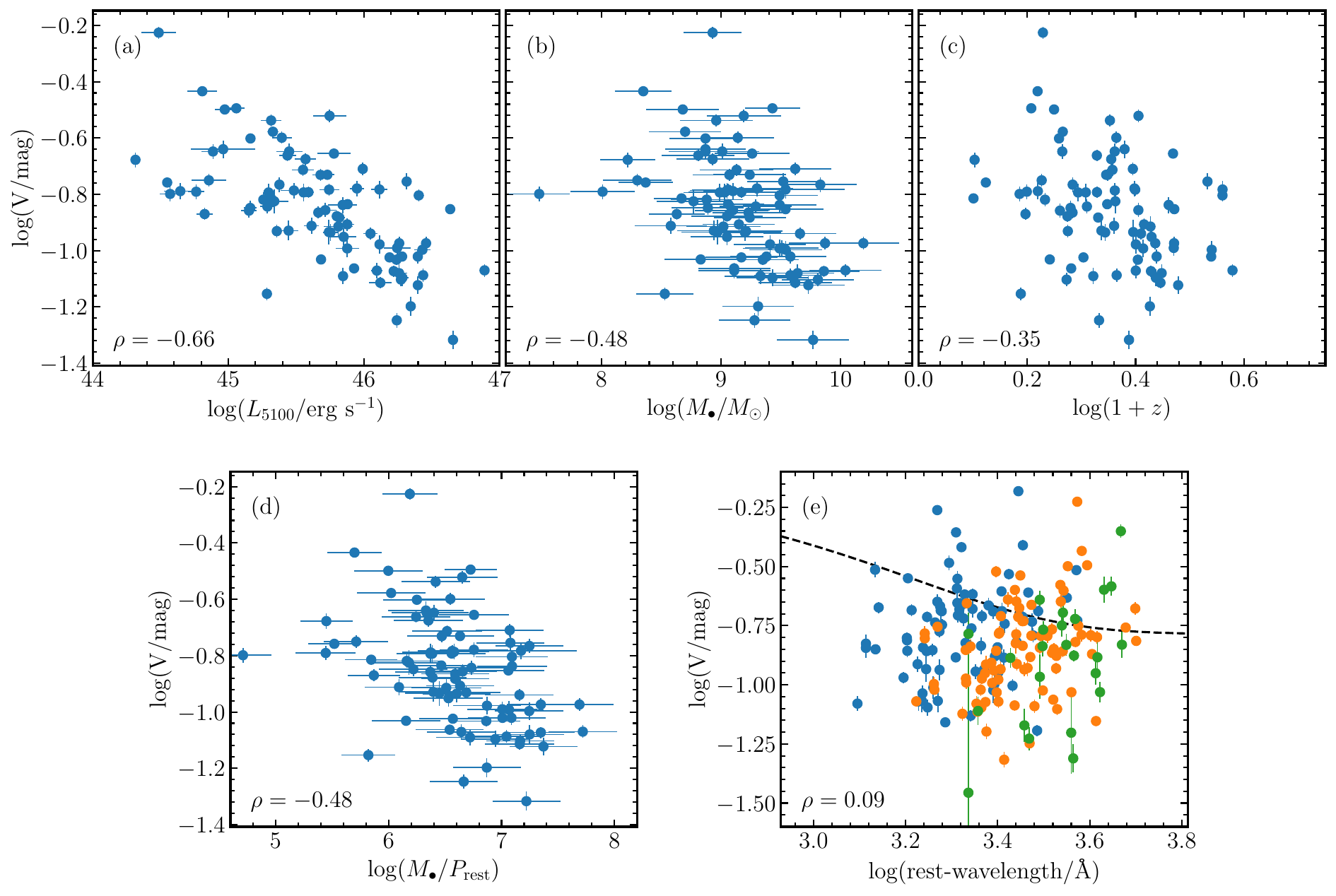}
	\caption{Variability amplitudes versus 5100~{\AA} luminosity, black hole mass, redshift, rest-frame wavelength and the ratio of black hole mass to period. The denotation $\rho$ marks Spearman’s rank correlation coefficient. In panels (a)-(d), the blue points are the variability amplitude in $r$-band. In panel (e), the blue, orange and green points are the variability amplitudes in $g$-, $r$- and $i$-band, respectively. The dashed line is from \citet{Vanden2004}.} 
	\label{fig_amp}
	\end{figure*}
 
    \subsection{Basic Properties of the Periodic Quasar Sample}
    \label{sec_properties}
	In appendix~\ref{sec_spec}, we detail how to estimate the 5100~{\AA} luminosity, black hole mass, and accretion rate of our sample from the SDSS archival spectra and tabulate the basic properties of our selected candidates in Tabel~\ref{table1}.
	
	Fig.~\ref{fig_hist} illustrates the distributions of the redshifts, $r$-band magnitudes, and periods from the GLSPs. Fig.~\ref{fig_properties} illustrates the distributions of the black hole mass, 5100~{\AA} luminosity, and accretion rate (see Appendix~\ref{sec_spec} for the definition).
	The redshift distribution peaks around $z\sim0.5-1.8$, generally coincident with the redshift distribution of the whole quasar sample. The magnitude distribution has a narrow range of about 17-20 mag and peaks around 19 mag, corresponding to a black hole mass of $\sim10^9M_\odot$ and luminosity of $\sim10^{45.5}~{\rm erg~s^{-1}}$.  
	The period ranges between 500 and 950 days and its distribution peaks around 850 days, but has a hard upper limit. This arises from the temporal baseline ($\sim$1500 days) that restricts the period to be less than $\sim$1000 days in the observed frame.
	
	Fig.~\ref{fig_properties} also plots the relations between the rest-frame period (from the GLSPs) and black hole mass, 5100~{\AA} luminosity, and accretion rate. There appear superficial anti-correlations in these relations, especially for the black hole mass and luminosity. A simple inspection indicates that these anti-correlations are caused by the redshift effect as follows. Both the observed periods and magnitudes are concentrated on a narrow range. As a result, the rest-frame period $P\propto(1+z)^{-1}$ and the luminosity $L_{5100}\propto D_L^2$ increases with $(1+z)$, where $D_L$ is the luminosity distance. The black hole mass can be deemed to be proportional to $L_{5100}$ if regardless of the accretion rate. These factors lead to the spurious anti-correlations between the rest-frame period and black hole mass/luminosity. 
	This can be further verified by a partial correlation analysis \citep{Kendall1979}. The correlation coefficient between $x$ and $y$ excluding the dependence on the third parameter $z$ is defined as
	\begin{equation}
	    r_{xy,z} = \frac{\rho_{xy}-\rho_{xz}\rho_{yz}}{\sqrt{1-\rho_{xz}^2}\sqrt{1-\rho_{yz}^2}},
	    \label{eqn_partial_corr}
	\end{equation}
	where $\rho_{xy}$, $\rho_{xz}$, and $\rho_{yz}$ are the Spearman’s rank correlation coefficients of $x$ versus $y$, $x$ versus $z$ and $y$ versus $z$, respectively. The Spearman’s rank correlation coefficients between $\log P_{\rm rest}$ and ($\log\bhm$, $\log L_{5100}$ and $\log\dot{\mathscr{M}}$) are $\rho=(-0.58,-0.68,-0.07)$  and their corresponding $p$-values are $p$=(4.5$\times10^{-9}$, 8.1$\times10^{-13}$, 0.5$\times10^{-1}$). Meanwhile, the coefficients  between ($\log P_{\rm rest}, \log\bhm$, $\log L_{5100}$ and $\log\dot{\mathscr{M}}$) and $z$ are $\rho=(-0.90,0.69,0.78,0.02)$, and their $p$-values are $p$=(1.5$\times10^{-30}$, 2.3$\times10^{-13}$, 4.2$\times10^{-19}$, 8.9$\times10^{-1}$). Thus, the partial correlation coefficients between $\log P_{\rm rest}$ and ($\log\bhm$, $\log L_{5100}$ and $\log\dot{\mathscr{M}}$) according to Equation~(\ref{eqn_partial_corr})  are $r=(0.11,0.09,-0.14)$ and their corresponding $p-$values for a null hypothesis test ($r_{xy,z}=0$) are $p = (0.67, 0.57, 0.79)$. Such low partial correlation coefficients indicate that there are no significant correlations between the rest-frame period and the black hole mass, 5100~{\AA} luminosity, and accretion rate. This is consistent with the results of \citet{Lu2016} based on the CRTS periodic quasar sample identified by \citet{Graham2015b}.
	
	\subsection{Variability Amplitudes}
        \label{sec_variability}
	To investigate the relation of intrinsic variability amplitudes versus luminosity, black hole mass, redshift, and rest wavelength, we measure the variability amplitudes by using the expression (\citealt{Vanden2004,Liu2019})
	\begin{equation}
	    V = \sqrt{\frac{\pi}{2}a_0^2-\epsilon^2},
	\end{equation}
	where $a_0$ is the amplitude of the best-fit sinusoidal function (see Section~\ref{sec_method}), $\epsilon^2 = \sum_{i=1}^{N}\sigma_i^2/N$, $\sigma_i$ is the magnitude uncertainties of the $i$-th point, and $N$ is the number of observations.
	
	Fig.~\ref{fig_amp}a-c plot the relation of $\log V$ in $r$-band with $\log L_{5100}$, $\log M_{\bullet}$ and $\log (1+z)$. There show anti-correlations in these relations, similar to previous studies on normal AGNs (e.g., \citealt{Kelly2009}). Regarding the relation between $\log V$ and redshift, however, several previous studies found a positive correlation (e.g., \citealt{Vanden2004}). It is unknown whether such a difference is intrinsic or caused by different selected quasar samples.
	
	In Fig.~\ref{fig_amp}d, we show the relation between $\log V$ and $\log (M_{\bullet}/P_{\rm rest})$. This plot is motivated as follows. The close binary supermassive black holes (SMBHs) have been proposed as a possible origination of periodicity in AGNs (e.g., \citealt{Bon2012,Farris2014,Graham2015a,DOrazio2015,LiYR2016,LiYR2019,Liu2019,Liao2021}). \cite{DOrazio2015} used the Doppler boosting of the emissions from the accretion disk surrounding the secondary black hole to explain periodic variations. In this scenario, the velocity of the secondary black hole in a circular orbit is 
	\begin{equation}
	    v_2 = \left(\frac{2\pi}{1+q}\right)\left(\frac{GM_{\bullet}}{4\pi^2P}\right)^{1/3},
	\end{equation}
    where $q=M_2/M_1\leq 1$ is the mass ratio, $M_{\bullet}=M_1+M_2$, $M_1$ and $M_2$ are the mass of the primary and secondary black hole, respectively, and $P$ is the orbital period. To the first-order approximation, the variability due to the Doppler boosting is 
    \begin{equation}
        \frac{\Delta F_\nu}{F_\nu} = (3-\alpha_\nu)\frac{v_2}{c}\cos{\phi}\sin{i},
    \end{equation}
    where $\Delta F_\nu$ is the flux at a specific frequency $\nu$, $\alpha_\nu$ is the spectral index, $\phi$ is the phase angle of the orbit, and $i$ is the inclination angle. For a large sample, we expect a relation $V\propto \Delta \log F_{\nu}\varpropto\log (M_{\bullet}/P_{\rm rest})$. However, as shown in Fig.\ref{fig_amp}d, $\log V$ has an anti-correlation with $\log (M_{\bullet}/P_{\rm rest})$. We notice that there are correlations of the variation amplitude, black hole mass, and period with redshift, which might cause a superficial anti-correlation between $\log V$ and $\log (M_{\bullet}/P_{\rm rest})$. Therefore, we perform a partial correlation analysis excluding the dependence on redshift as in Section~\ref{sec_properties}. We find a partial correlation coefficient of $r=-0.35$, which is marginal and possibly indicates that the anti-correlation is not intrinsic. 

    \begin{figure}
	\centering
	\includegraphics[width=0.42\textwidth]{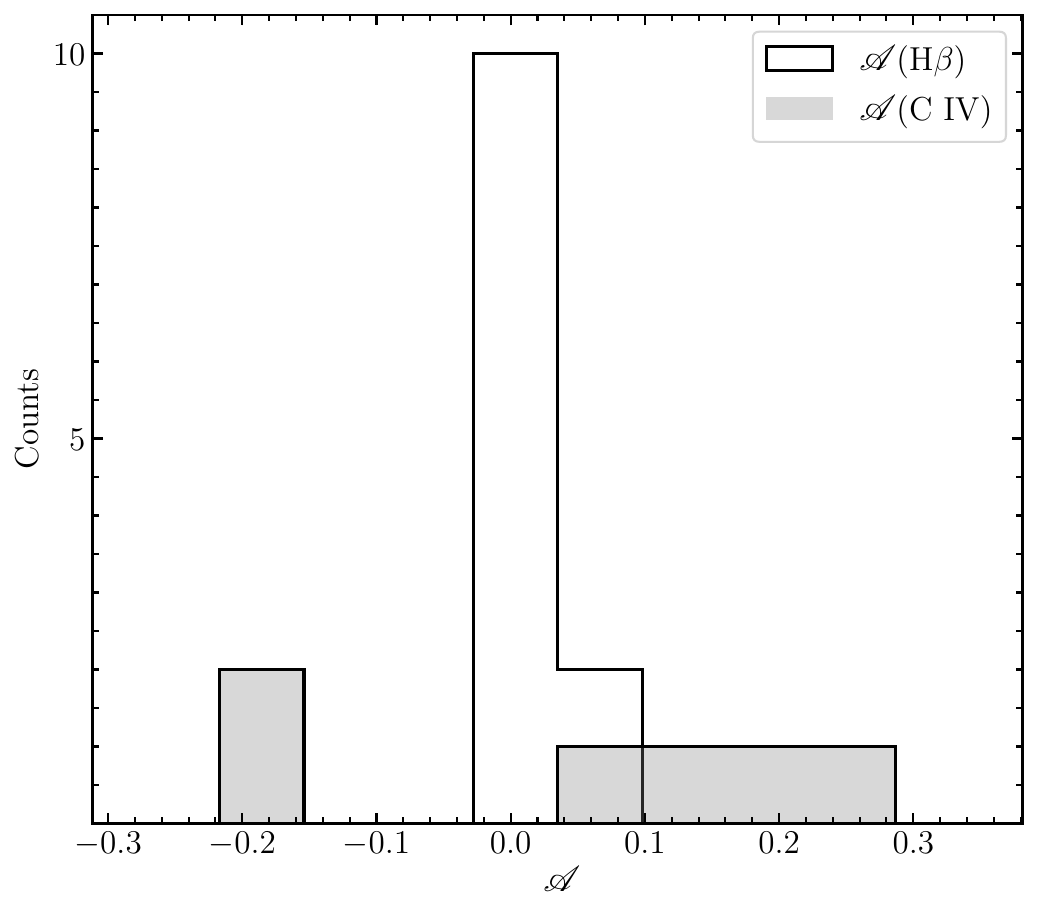}
	\caption{The distributions of $\mathscr{A}$ for the H$\beta$ and {\civ} lines of our periodic quasar sample.}
	\label{fig_histA}
	\end{figure}
 
    As mentioned above, the ZTF has three custom filters $gri$. Their effective wavelengths are $\lambda_{\rm eff}(g)=4722.74$~\AA, $\lambda_{\rm eff}(r)=6339.61$~\AA, $\lambda_{\rm eff}(i)=7886.13$~\AA, respectively. In the above analysis, we by default use the $r$-band data. We can also calculate the variability amplitudes for the data of the other two bands. As such, we can obtain the relation between the variability amplitude with the rest-frame wavelength. We use the light curves with more than 40 binned points for $g$-band and more than 20 binned points for $i$-band (because the $i$-band generally has poor sampling). Fig.~\ref{fig_amp}e illustrates the dependence of the variability amplitudes on rest-frame wavelength, which is calculated by dividing the effective wavelength by the corresponding $(1+z)$. The variability amplitude of our sample has no obvious declining trend with rest-frame wavelength. This is opposite to the results of~\citet{Vanden2004} and~\cite{Liu2019}, which, however, found the variability amplitude decreases with rest-frame wavelength as $V\propto\exp{(-\lambda/988{\text{\AA}}})$. A possible cause of this discrepancy may be due to the anti-correlation between variability amplitudes and redshift in our sample (see Fig.~\ref{fig_amp}c).
    
    With future development of photometric sky surveys, such as the ongoing ZTF and the upcoming Vera C. Rubin Observatory’s Legacy Survey of Space and Time (\citealt{Ivezic2019}), a larger sample of periodic quasar candidates with much longer temporal baselines and greater diversity in period and redshift is highly worthwhile to testify the above correlations.
    
    \subsection{The Asymmetry of Broad Line Profiles}
    To quantity the asymmetry of the broad emission lines, we adopt the dimensionless asymmetry parameter (\citealt{Brotherton1996,Du2018})
    \begin{equation}
        \mathscr{A} = \frac{\lambda_c(3/4)-\lambda_c(1/4)}{\Delta \lambda(1/2)},
    \end{equation}
    where $\lambda_c(3/4)$ and $\lambda_c(1/4)$ are the wavelengths at the 3/4 and 1/4 of the peak height, and $\Delta \lambda(1/2)$ is full with at half maximum (FWHM). The line profile has a blue asymmetry for $\mathscr{A}$ > 0 and a red asymmetry for $\mathscr{A}$ < 0. We calculate the asymmetry parameters of the prominent H$\beta$ and {\civ} profiles obtained by the spectral decomposition described in Appendix~\ref{sec_spec}. For the prominent {\mgii} line, we find that it is sufficient to use one single Gaussian in the spectral decomposition so that the resulting asymmetry parameters $\mathscr{A}=0$. In Fig.~\ref{fig_histA}, we plot the distributions of $\mathscr{A}$ for the H$\beta$ and {\civ} lines of our sample. Most of the H$\beta$ profiles are roughly symmetric ($\mathscr{A}\sim0$) while the {\civ} profiles are either red or blue asymmetric. 
    
    In the scenario of binary SMBHs, the broad emission lines are expected to be asymmetric due to the orbital motion of the black holes that cause velocity shifts to the lines emitted from the broad-line regions (BLRs) surrounding each black hole (e.g.,  \citealt{Shen2010,Eracleous2012,Bon2012,Bon2016,LiYR2016,LiYR2019}). However, if the orbital separation is smaller than the BLR size, the binary shares a mutual circumbinary BLR and the influences of the orbital motion might be minimized. The generated broad emission line profiles will be similar to those of normal AGNs. We estimate the orbital separations of our sample using Kepler's law
    \begin{equation}
    \frac{A}{R_{\rm g}} = 170.9 \left[\left(\frac{M_\bullet}{10^9M_\odot}\right)^{-1}\left(\frac{P}{800~\rm day}\right)\right]^{2/3},
    \end{equation}
    where $R_{\rm g}$ is the gravitational radius. We estimate the H$\beta$ BLR sizes using the well-established size-luminosity relation (see Appendix~\ref{sec_spec}). As shown in Fig.~\ref{fig_SepVsR_BLR}, the orbital separations are overall smaller than the BLR sizes.
    
    On the other hand, we note that asymmetric emission lines can also be produced from the BLR of a normal AGN with complicated geometry and dynamics, such as inflows and/or outflows (e.g., \citealt{Denney2009,Grier2013,Fausnaugh2018,DeRosa2018,Bao2022,U2022,Lu2022,Chen2023}), spiral arms (e.g., \citealt{Gilbert1999,Storchi-Bergmann2003,Storchi-Bergmann2017,Wang2022, Du2023}), and partial dust obscuration (e.g., \citealt{Gaskell2018,Panda2022}). In this sense, only using one-epoch line asymmetry, we cannot distinguish binary black hole systems from single black holes. However, long-term spectroscopic monitoring of those periodic candidates can yield time variations of the line asymmetry, shedding more light on testifying binary black hole systems.

    \begin{figure}
    \centering
    \includegraphics[width=0.43\textwidth]{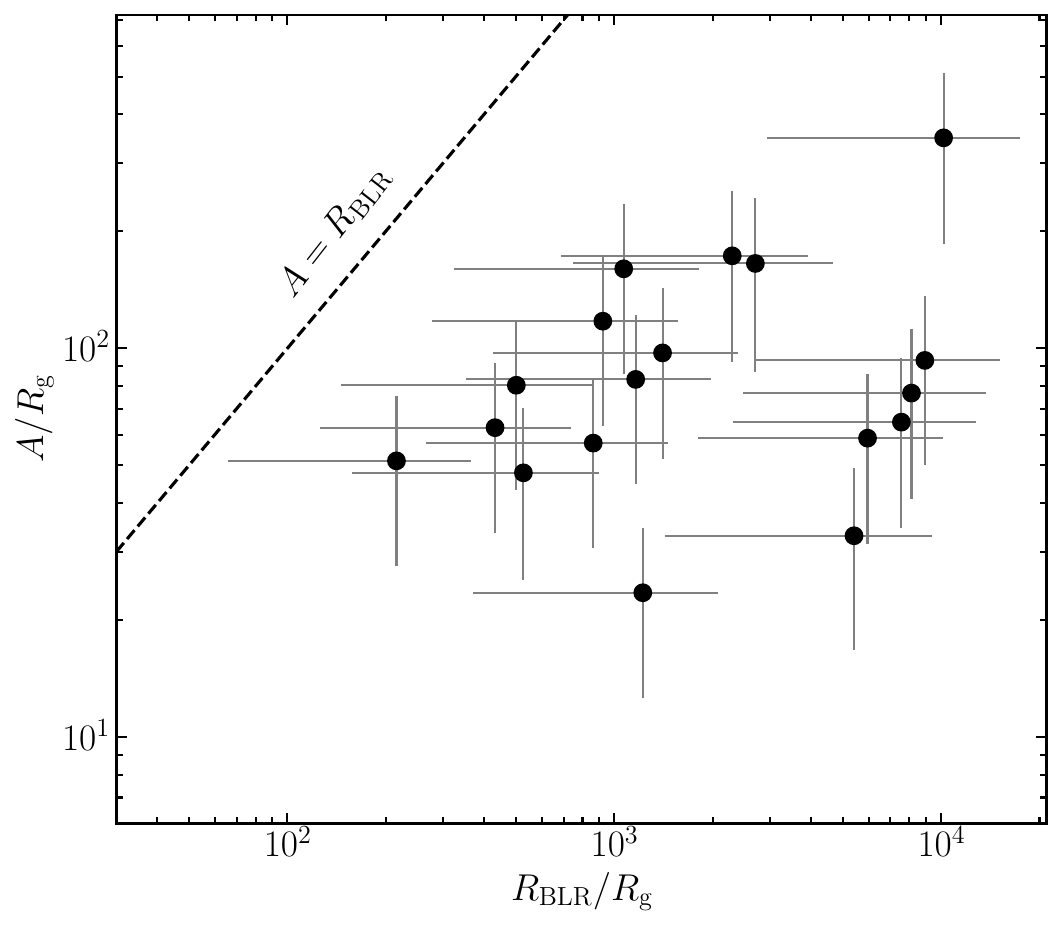}
    \caption{Comparison of the orbital separations versus the H$\beta$ BLR sizes by assuming that the periodicity arises from the orbital motion of supermassive black hole binaries and the BLR size follows the size-luminosity relation.}
    \label{fig_SepVsR_BLR}
    \end{figure}

    \begin{figure*}
        \centering
        \includegraphics[width=0.65\textwidth]{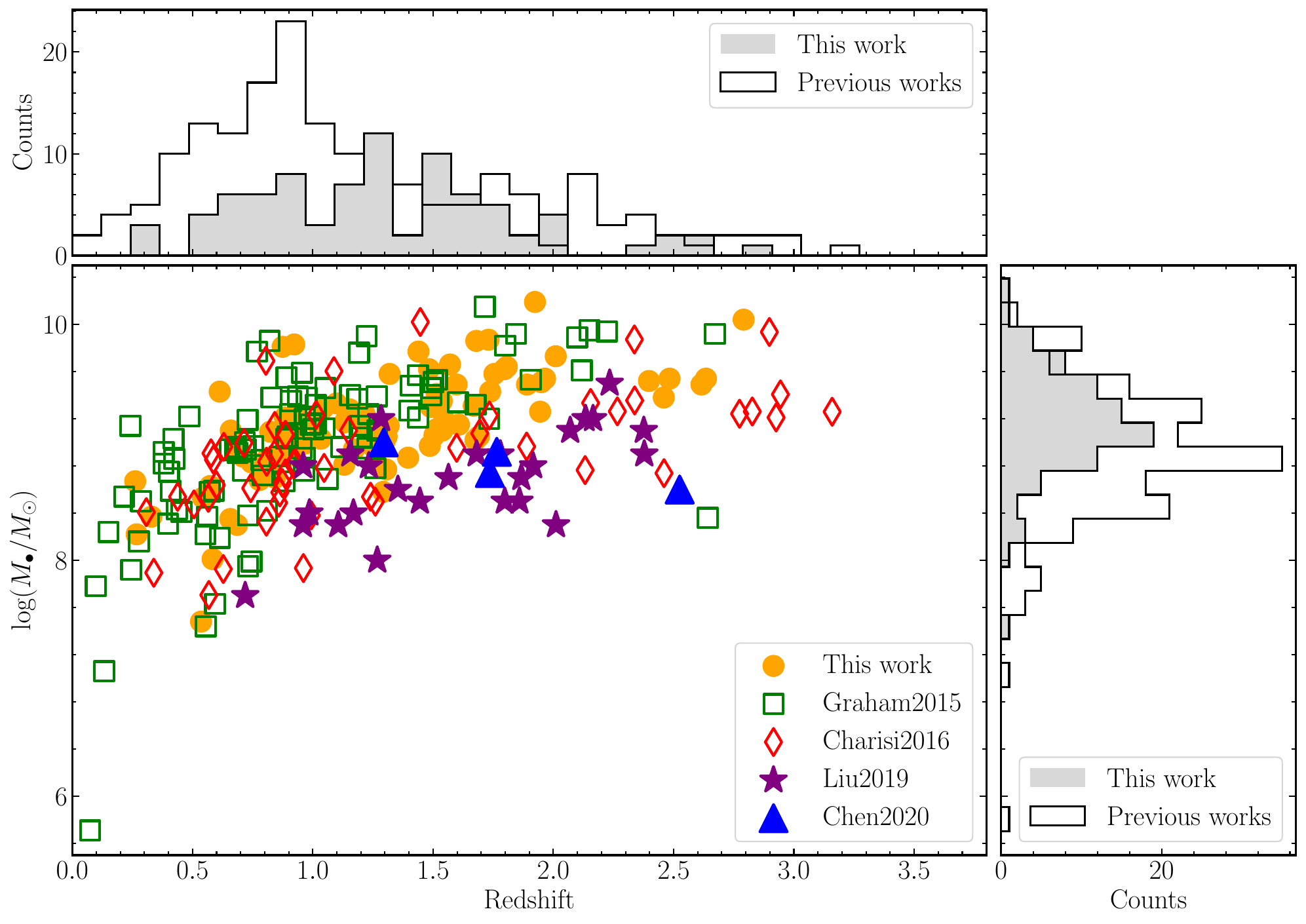}
        \caption{The distribution of the redshift and black hole mass for the periodic candidates found in this work and previous works (\citealt{Graham2015b,Charisi2016,Liu2019,Chen2020}).}
        \label{fig_compare}
    \end{figure*}
    %
	
    \subsection{Comparison with the Previous Works}
	With systematic searches over survey data, previously there were 111 objects identified from CRTS \citep{Graham2015b}, 50 from PTF \citep{Charisi2016}, 26 from PS1 MDS \citep{Liu2019}, and 4 from the combination of the DES and SDSS \citep{Chen2020}. 
    Fig.~\ref{fig_compare} compiles the black hole mass of the previous samples together with our sample and plots the relationship between the black hole mass with redshift. The overall redshift distributions of the black hole mass of those samples are similar.
	
	The CRTS sample of \cite{Graham2015b} was based on the combination of the Million Quasars (MQ) v3.7 catalogue and SDSS DR12 quasar catalog. \citet{Charisi2016} used the Half MQ catalog (a subsample of the MQ Catalogue v4.4) and SDSS DR12 quasar catalog as the input quasar catalogs. As a comparison, we use SDSS DR14 quasar catalog and the \citeauthor{Veron2010} catalog.
	Therefore, we expect that there are significant overlaps among these three input catalogs. However, we cross-match our periodic sample with the samples of \cite{Graham2015b} and \cite{Charisi2016}, 
    and do not find matching candidates. 
	The possible reasons that there are no matching candidates among the three samples are as follows: 1) the temporal baseline is crucial and different temporal baselines may lead to different quasars identified. For example, most of the candidates from \cite{Graham2015b} have observed periods larger than 1400 days, the candidates from \cite{Charisi2016} have observed periods less than 500 days, and our sample has observed periods in the range from 500 to 950 days. 
     2) The period {and variation amplitude} might change with time (e.g., \citealt{Jiang2022}). 3) The periodicity might only appear at some epochs and disappear at other epochs (e.g., \citealt{ONeill2022}). {The latter two behaviours can be explained once the variability is not strictly periodic, but quasi-periodic in the sense that the corresponding PSD shows a broad peak on top of a red-noise-like shape (see Section~\ref{sec_periodicity_verify} for more discussions).}

    \begin{figure*}
        \centering
        \includegraphics[width=0.90\textwidth]{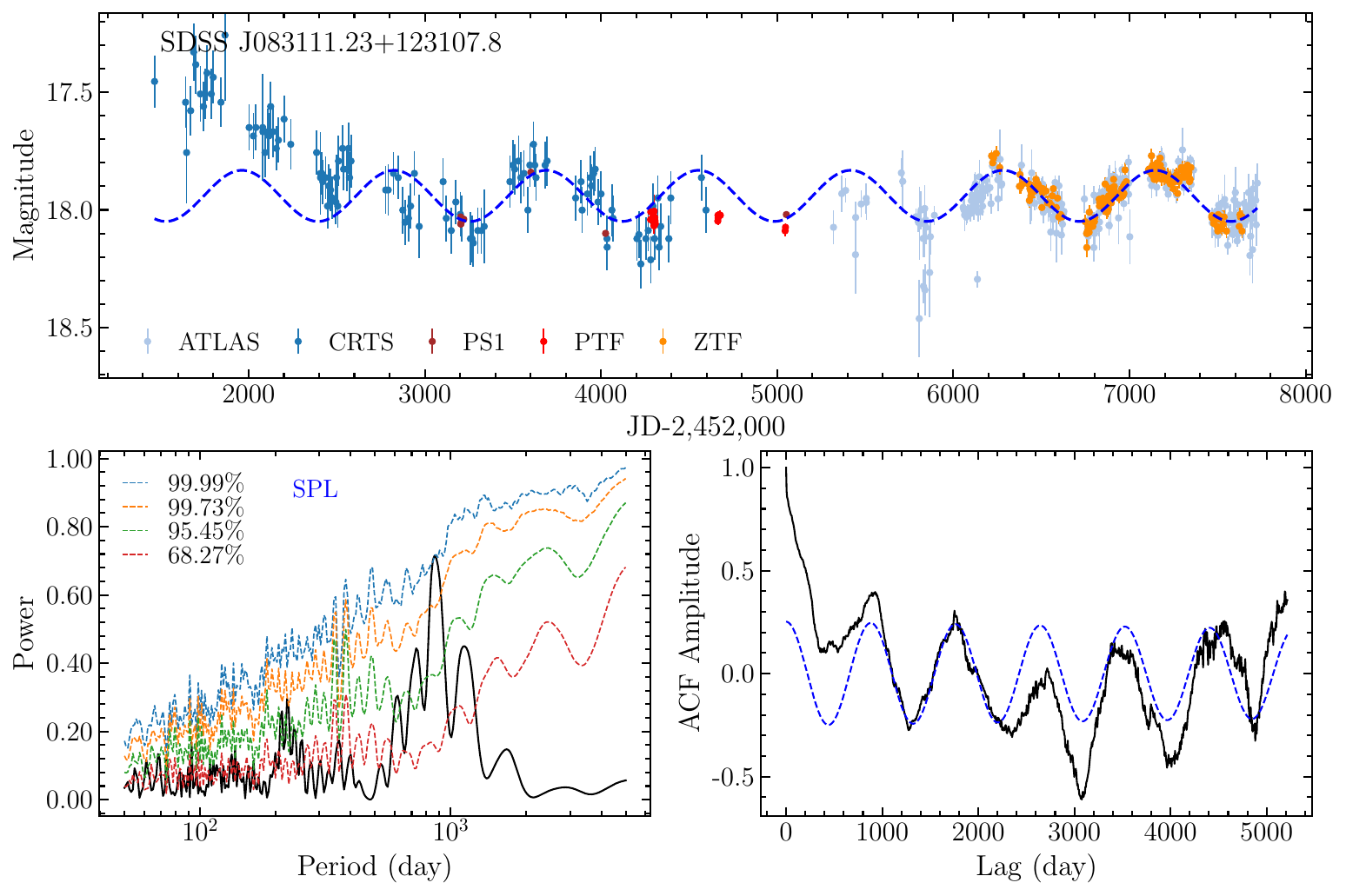}
        \includegraphics[width=0.90\textwidth]{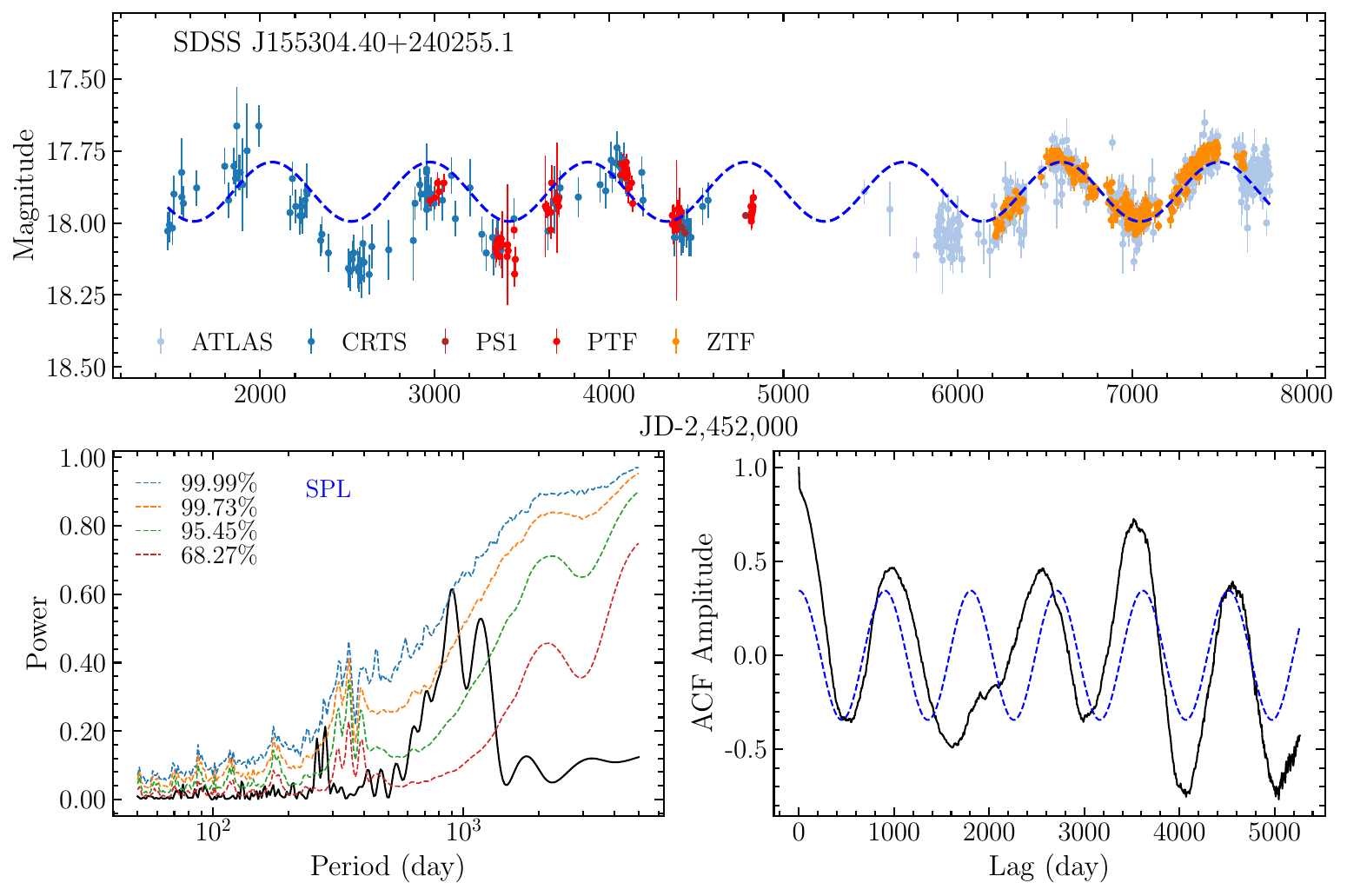}
        \caption{Three periodic quasar candidates selected from extended light curves. For each candidate, the upper panels show the synthetic light curves from the CRTS (blue), PS1 (brown), PTF (red), ATLAS (light blue), and ZTF (orange) archival survey data. The blue dash line shows the best sinusoidal fit. The bottom left panel shows the GLSP and the significance levels in terms of the SPL red-noise model. The bottom right panel shows the ACF of the extended light curve (the black solid line) and the best-fit using an exponentially decaying cosine function (the blue dashed line).}
        \label{fig_lc_extend}
    \end{figure*}

    \begin{figure*}
        \centering
        \includegraphics[width=0.90\textwidth]{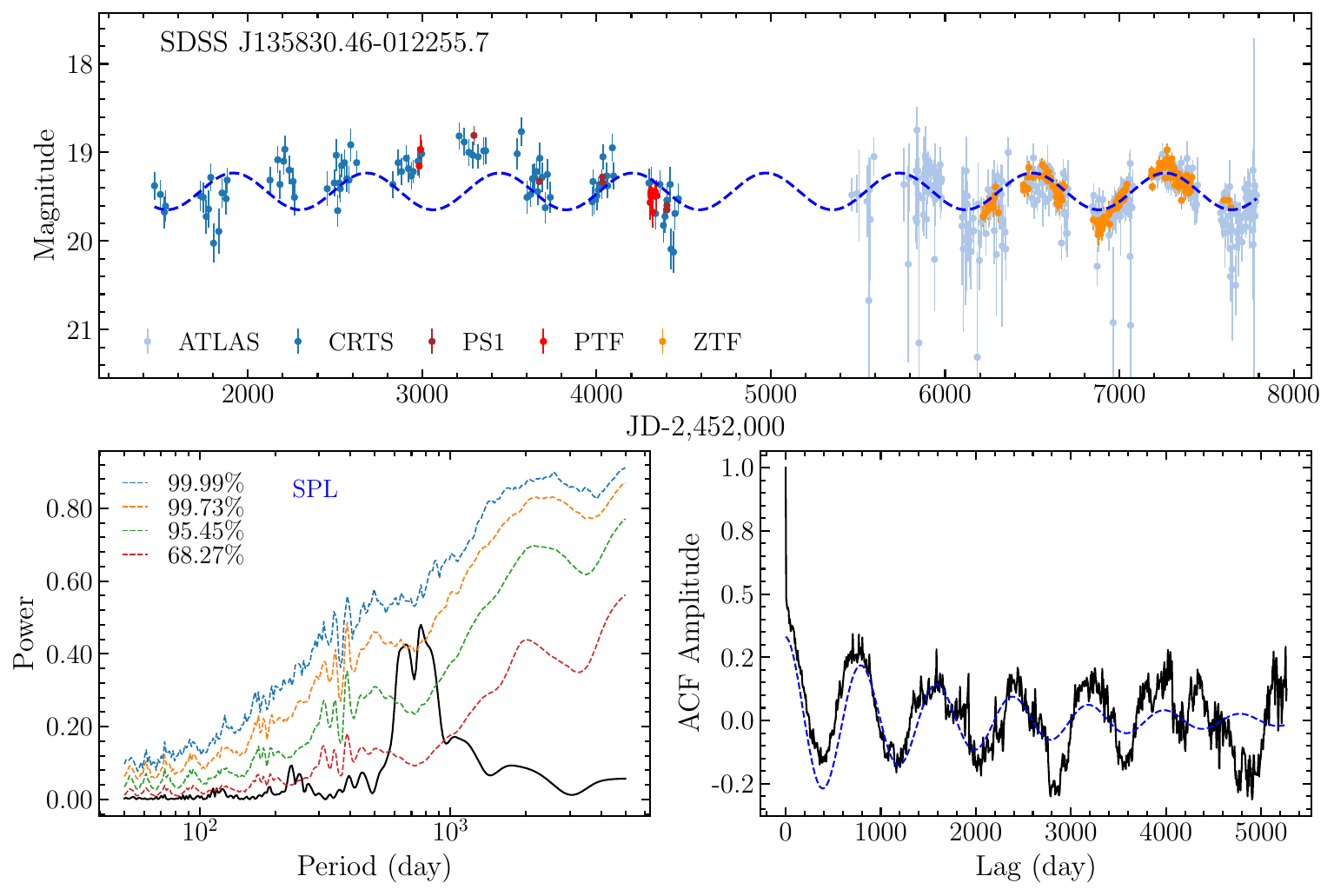}
        \addtocounter{figure}{-1}
        \caption{(Continued).}
        \label{fig_lc_extend}
    \end{figure*}

    \section{Discussions}
    \label{sec_discussion}

    \subsection{The Look-elsewhere Effect from Searching over an AGN Sample}
    \label{sec_global_fap} 
    In Section~\ref{sec_fap}, we have estimated the FAP for an individual AGN, which includes the look-elsewhere effect arising from searches over a broad period range. Another type of look-elsewhere effect is the false positive when globally searching over an AGN sample due to random fluctuations. To estimate this kind of ``global FAP'', we perform the following simulations. We generate 143,700 mock light curves with the same method in Appendix~\ref{sec_psd_test} by assuming that the AGN variability is purely stochastic and modeled by SPL.
    The SPL slopes are randomly assigned from a range of 1.5-3.5, based on the slope distribution of 184 candidates (see Figure~\ref{fig_FAP_Comparison}). We then search over these mock light curves to identify ``fake'' periodic candidates using exactly the same procedures employed for the identification of the 86 candidates. We select out a total of 51 candidates, implying a global FAP $\sim3.6\times10^{-4}$. {\it This result indicates that in a statistical sense, most of our 86 periodic candidates might be normal AGNs and their spurious periodic variations are caused by red-noise. }
    
    In light of the temporal limitations of ZTF data, additional data will be helpful to reduce the global FAP and verify/falsify the periodicity. In next section, we extend the temporal baselines of ZTF light curves by compiling other time-domain survey data.    
    
    \subsection{Extended Light Curves}
    \label{sec_extendedlc}
    \subsubsection{Photometric Data Intercalibration}
\begin{figure*}
        \centering
        \includegraphics[width=0.98\textwidth]{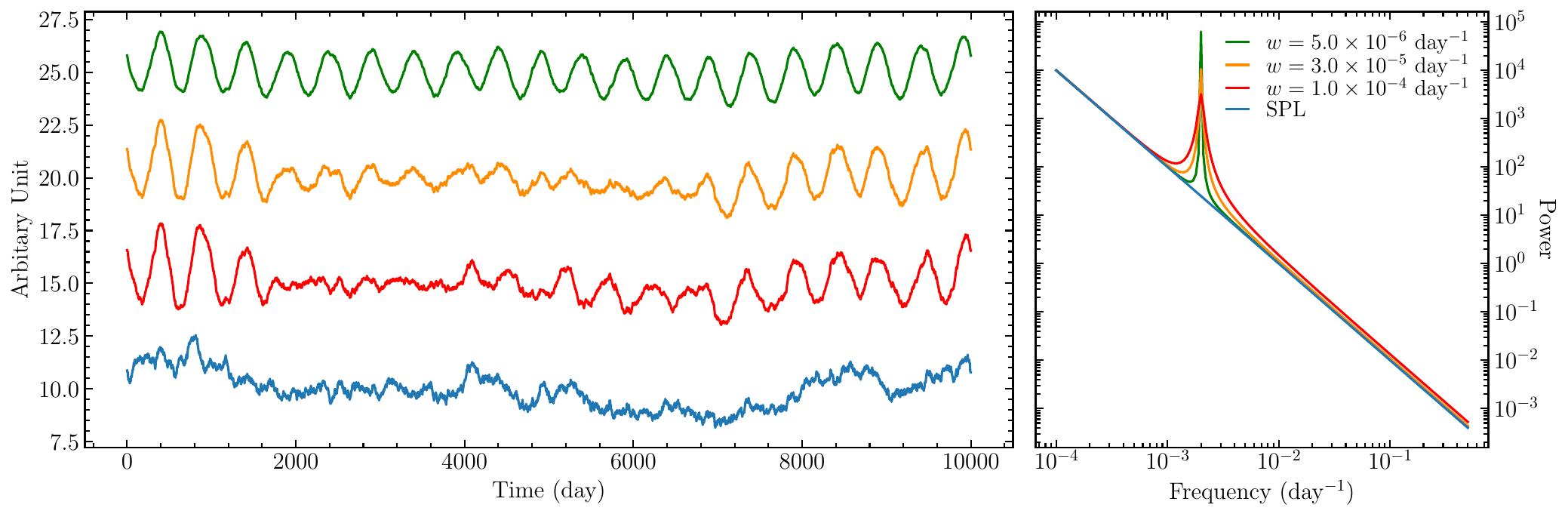}
        \caption{Mock light curves (left) and the corresponding PSDs (right). The colors of light curves match those of PSDs. In the right panel, the solid blue line shows the pure SPL model with a form of $P_{\rm SPL}=10^{-4}\times f^{-2}$, where $f$ is the frequency. 
        The orange, green and red-solid lines show composite PSDs of the SPL plus a Lorentzian function $P_{\rm Lor}=w/{\pi[w^2+(f-f_c)^2]}$ with different widths $w$ and a fixed center frequency $f_c=2\times10^{-3}$ day$^{-1}$.  
        }
        \label{fig_mock_periodic_lc}
    \end{figure*}

    {By incorporating archival survey data from CRTS, PTF, PS1, and ATLAS (short for Asteroid Terrestrial-impact Last Alert System), we extend the light curves of the 86 periodic candidates with high statistical significance to further verify the periodicity. Here, we utilize $r$-band data of PTF and PS1 and $c$-band data of ATLAS. Given that these photometric data were obtained with different instruments, filters, and reduction methods, it is imperative to perform intercalibration between them. For the data obtained from ATLAS, the monitoring period temporally overlapped with that of ZTF. We perform intercalibration using the Bayesian package {\tt PyCALI}\footnote{\url{https://github.com/LiyrAstroph/PyCALI}.} (\citealt{Li2014}) by adopting the ZTF light curve as a reference. As for the data from CRTS, PTF and PS1, we first adopt the $r$-band light curve from PS1 as the reference and align other light curves with this reference. We then convolve the SDSS spectrum with the PS1/ZTF filter transmissions to obtain synthetic magnitudes and calculate the magnitude difference between the two photometric systems.
    Finally, we add this magnitude difference to the aligned data of CRTS, PTF, and PS1 to keep consistent with the ZTF data.
    In Appendix~\ref{sec_lc}, we show the intercalibrated long-term light curves of the 86 periodic candidates.

    \subsubsection{Verifying the Periodicity}
    \label{sec_periodicity_verify}
    With the extended light curves, we can verify the periodicity using the same methods described in Section~\ref{sec_method}\footnote{We did not start with extended data to identify periodic samples in Section~2.2 because of two main reasons. Firstly, 
    The ATLAS database provides on-the-fly reduced photometry and acquiring 143,700 objects is unrealistic presently. Secondly, accurately intercalibrating fluxes for all archived data poses a challenge for some objects due to the lack of SDSS spectra and/or there being no time overlap between CRTS, PTF, PS1, ATLAS, and ZTF. }. Unfortunately, we find that no candidate satisfies the criterion of $\xi>4.0$. This indicates that the light curves of the periodic candidates may not rigorously adhere to the expected sinusoidal form due to contamination of red-noises (see below), the influence of poor photometric uncertainties, or other unknown factors. With this consideration, we reidentify the periodic quasars by discarding the criterion of $\xi>4.0$ but still requiring the peak power in the GLSP exceeding 99.73\% (in terms of using the SPL model as the null hypothesis), $\left|1-P_{\rm ACF, Extend}/P_{\rm GLS, Extend}\right|\textless 0.1$, and $\lambda<10^{-3}$ day$^{-1}$. Meanwhile, we imposed an additional criterion that the difference between the GLSP periods from extended and ZTF light curves should be within 10 percent.

    Finally, three candidates (i.e., SDSS J083111, SDSS J155304 and SDSS J135830; see Figure~\ref{fig_lc_extend}) with 6-8 cycles of periods are identified using the extended data. In particular, the candidates SDSS J083111 and SDSS J155304 reach a significance level of 99.99\%. Notably, it seems that the periodic variation in the light curve of SDSS J083111 started to appear around JD 2,454,000 (see the upper panel of Figure~\ref{fig_lc_extend}), after a rapid declining trend that apparently deviates from the expected periodic variability. Such behavior is reminiscent of the candidate PKS~2123-021 reported by \cite{ONeill2022}, which exhibits periodic variability only at certain epochs (see Fig. 1 therein).

    For the remaining 83 periodic candidates that do not satisfy the criteria, there are two possibilities. First, the periodic variations might be spurious and purely originate from red-noises as discussed in Section~\ref{sec_global_fap}.
    However, given the sparse sampling and/or large photometric uncertainties of the extended data, more monitoring data are required to finally confirm this point. Alternatively, as mentioned above, the observed behaviours reflect that the variability is quasi-periodic instead of strictly periodic. Apparently, the significance level of the periodicity under the present criteria will be diminished if the period varies over time~(e.g., \citealt{Jiang2022}) or is only present during certain epochs (e.g., \citealt{ONeill2022}). This makes it challenging to identify true periodic quasars and estimate the significance of the periodicity. 

    For the sake of illustration, in Fig.~\ref{fig_mock_periodic_lc} we artificially generate mock light curves using PSDs composed of a SPL with a slope of $\beta=2$ and a Lorentzian function. As can be seen, in the cases of mild Lorentzian widths, the generated light curves show sinusoidal patterns at some epochs but seemingly stochastic patterns elsewhere. Also, it appears that the amplitudes of the sinusoidal patterns vary with time. 
    As expected, the generated periodic signal becomes more visible when the Lorentzian width decreases. Besides the Lorentzian width, the underlying SPL also introduces extra stochasticity to the (quasi-)periodic signals.
    }

    \subsection{Possible Explanations of Periodic Variability}
    {
    In this section, we present a brief discussion on the explanations of the periodicity based on the three candidates identified with extended light curves by assuming that the periodicity is real (see also \citealt{Graham2015b} and \citealt{Dotti2023}).
    }

    \begin{table}
        \renewcommand{\arraystretch}{1.2}
        \setlength{\tabcolsep}{4pt}
        \caption{Time delays with the ICCF method and phase differences of the best sinusoidal fits for the $g$-, $r$- and $i$-bands light curves of three candidates. The $i$-band light curves of SDSS J083111 and SDSS J135830 have too few data points to obtain reliable time delays and phase differences. \label{table_timelags}}
        \begin{tabular}{lcccc}
        \hline
        ID & $\tau_{gr}$ & $\tau_{gi}$ &  $\Delta \phi_{gr}$ &  $\Delta \phi_{gi}$\\
         &  (day) &  (day) & (day) & (day)\\
        \hline
        SDSS J083111 & $-8.1\pm10.7$ & --- & $-13.7\pm4.7$ & ---\\
        SDSS J155304 & $13.5\pm5.9$ & $35.3\pm10.9$  & $11.2\pm4.0$ & $39.0\pm9.6$\\
        SDSS J135830 & $-8.2\pm12.5$ & ---  & $0.6\pm8.5$ & --- \\
        \hline
        \end{tabular}
    \end{table} 
    
    \subsubsection{Doppler Boosting of Supermassive Black Hole Binaries}
    \label{sec_doppler_boosting}
    {
    This scenario was proposed by \cite{DOrazio2015}, in which the periodicity arises from the Doppler-boosted emissions of the accretion disk surrounding the secondary black hole.
    In Section~\ref{sec_variability}, we have tested this scenario by checking on the correlation between variability amplitudes and $\log (M_{\bullet}/P_{\rm rest})$ for the 86 candidates. We do not find a significant, expected positive correlation by excluding the selection bias.

    Another testable point regarding the Doppler boosting scenario is that the resulting relative flux and thus magnitude variations are expected to be wavelength-independent and simultaneous (i.e., no time delay) among different bands. To test this point, we measure time delays among the $g$-, $r$- and $i$-bands ZTF light curves using two methods. The first method relies on calculating the ICCF, from which the time delay is determined as the ICCF centroid above 80\% of the peak value.
    We denote $\tau_{gr}$ and $\tau_{gi}$ as time delays of the $r$- or $i$-bands light curves relative to the $g$-band one, respectively. 
    To reduce the influence of short-timescale variability (ascribed to the accretion disk emissions), we smooth the light curves using a median filter of 7 points prior to ICCF analysis. An alternative method is calculating phase differences among the best sinusoidal fits (with the same fixed period) for the three-band light curves.  We define $\Delta \phi_{gr}$ =$P(\phi_r-\phi_g)/2\pi$ and $\Delta \phi_{gi}$ =$P(\phi_i-\phi_g)/2\pi$, where $\phi_g$, $\phi_r$ and $\phi_i$ are the best-fit phases, and $P$ is the period. 
    
    We list time delays and phase differences (in the observed frame) in Table~\ref{table_timelags}. The obtained time delays and phase differences are consistent with each other within uncertainties. For the candidate SDSS J155304, there are significant time delays between $g$- and $r$/$i$-bands, suggesting that the periodicity might not be caused by the Doppler boosting. For the other two candidates, the time delays and phase differences have large uncertainties and therefore no meaningful constraints can be made.
    }

    \subsubsection{Periodic Accretion of Supermassive Black Hole Binaries}
    {
    The orbital motion of supermassive black hole binaries can produce periodic modulation to mass accretion onto each black hole, giving rise to periodic emissions of the accretion disks.  Previous hydrodynamic simulations of circumbinary accretion discs predicted that the periodic pattern would exhibit a sawtooth shape rather than a sinusoidal form (e.g., \citealt{Farris2014}). The quality of the present light curves does not allow us to reliably distinguish the shapes of periodic variations. Nonetheless, in the case of SDSS J155304, the detection of significant inter-band time delays can be explained under this scenario.
    }

    \subsubsection{Other Explanations}
    {
    The observed optical flux variations in quasars may be a result of the combination of thermal emissions from the accretion disk and non-thermal emissions from the Doppler-boosted radio jet. The periodic variations arise from the jet precession, which might be driven by the orbital motion of a binary SMBH system or an internally rotating jet flow of a single black hole system. However, after cross-matching with the sample of radio-loud AGNs from~\cite{Best2005}, we find that none of the three candidates were included in this sample. This rules out the precessing jet scenario.
    Alternatively, in addition to radio jets, periodic variations could also be attributed to the precession of a warped accretion disk surrounding a single black hole, which results in the obscuration of the emission region \citep{Martin2007}. However, as discussed by \cite{Graham2015a} and \cite{Chen2020}, for a $10^{8}M_\odot$ SMBH, the precessing timescale is on the order of several tens of years. This is much longer than the periods ($\sim$2.5 years) of the three periodic quasar candidates, whose SMBH masses are in a range of $10^{8.3-9.0}M_{\odot}$.
    
    \subsection{Caveats and Future Improvements}
    \subsubsection{Caveats in the Methodology of Periodicity Identification}
    {There are two caveats in the method we adopted to identify candidates in Section~\ref{sec_method}. 
    
    Firstly, we initially select the candidates by using the best-fit sinusoidal fit with the period fixed to $P_{\rm GLS}$, which corresponds to the largest periodogram peak. In this step, we assume that the periodic signal is solely affected by white noises, namely, the noise power is constant across frequency. However, it is important to note that red-noise power decreases with frequency, as shown in the bottom left panels of Figure~\ref{fig_lc}. Consequently, our approach may result in the exclusion of some periodic candidates with short periods. A possible method to surmount this issue is to pre-select the period corresponding to the periodogram peaks that exhibit the highest significance in light of the red-noise background (the peak power in GLSP reaches 3$\sigma$ significance level at least), and then perform the same subsequent procedures as in Section~\ref{sec_method} such as sinusoidal function fitting. The red-noise background can be constructed by assuming the PSD follows some specific shapes (e.g.,  either DRW or SPL; see details in Section~\ref{sec_significance_freq}). In Section~\ref{sec_competition}, the comparison between DRW and SPL models reveals that the latter is more favorable for our quasar sample. However, due to the irregular sampling and seasonal gaps, the SPL parameters cannot be directly recovered by fitting the PSD (see Appendix~\ref{sec_psd_test}). Although the \texttt{RECON} framework offers an alternative method to obtain reliable PSD parameters, it is not feasible to construct the red-noise background for each of 143,700 quasars because of the prohibitive computational time required for running \texttt{RECON}.
    
    Here, we carry out a simple test to obtain a quantitative estimation of the number of candidates with shorter periods that may be missed when using the method in Section~\ref{sec_method}. We adopt the DRW process instead of the SPL model  and construct the red-noise background for each object in our parent sample. The subsequent steps and selection criteria fully follow those in Section~\ref{sec_method}.
    We finally identify 125 candidates covering a period range of 400-900 days, in which 90 objects overlap with our sample of 184 candidates identified in Section~\ref{sec_method}. These results imply that there are no serious missing identifications of short-period candidates in Section~\ref{sec_method}.
     }
     
     Secondly, the adopted criteria in the present work are inclined to select out candidates with sinusoidal variations. 
     Specifically, the limit on S/N ratio $\xi>4$ is sensitive to the amplitude of the sinusoidal variations. In addition, for the ACF selection script, the exponentially decaying cosine function 
     in Equation~(\ref{eqn_acf}) indeed corresponds to a noise-modulated sinusoidal signal with a form of (\citealt{Jung1993})
     \begin{equation}
      f(t) \propto \sin[2\pi t/P_{\rm ACF} + \phi + w(t)],
     \end{equation}
     where $\phi$ is the phase and $w(t)$ is the Wiener noise with a zero mean and correlation function of $\langle w(t_1)w(t_2)\rangle = (2\lambda) {\rm min}(t_1, t_2)$. Here, $\lambda^{-1}$ represents the coherence time of the phase fluctuations in the sinusoidal signal. We require $\lambda < 10^{-3}~{\rm day}^{-1}$, meaning that the coherence time is as long as 1000 days, comparable to the time baselines of ZTF data. As a result, this selection criterion also tends to identify sinusoidal variations, although not as sensitive as the S/N limit. 

    \subsubsection{Future Improvements for the Searching Method of Periodicity}
     Considering that realistic variations might deviate from a sinusoidal form
     (e.g., \citealt{Farris2014}) and also be quasi-periodic with a broad distribution of the period (see Fig.~\ref{fig_mock_periodic_lc}), it is necessary to devise new effective identification methods in future works. A possible starting point might be resorting to PSD fitting. However, since AGN light curves usually have highly irregular sampling and significant inhomogeneity in observing instruments and data reduction, the traditional direct PSD fitting as in X-ray analysis (e.g., \citealt{Uttley2002}) is no longer applicable.
    \cite{Li2018} proposed a forward approach to recover any given shapes of PSDs from irregularly sampled light curves with measurement noises. This approach can be easily extended to the case of a quasi-periodic PSD on top of a red-noise PSD. The Bayes factor can be employed to distinguish a quasi-periodic PSD from a pure red-noise PSD.

    We note that since a quasi-periodic PSD has distributed frequencies (and thereby periods), reliable identification of such signals would be challenging and require longer temporal baselines and better sampling rates. We expect that as the ZTF data continuously accumulates and other time-domain surveys come into operation in the near future (such as the Rubin-Large Synoptic Survey Telescope; \citealt{Ivezic2019}),  there will be a sizable sample of statistically significant quasi-periodic quasar candidates finally identified.
     }
    
    \section{Summary}
    \label{sec_summary}
    Based on the quasar catalogs of SDSS DR14 and \cite{Veron2010}, we systematically search for quasars with periodic variations over the ZTF archival light curve data. As a first step, we employ the GLSP and ACF methods to make a primitive selection. We restrict that there are at least 1.5 cycles of periods and sufficiently strong periodic signals in the light curves (see Section~\ref{sec_method}). Also, we require the difference between the obtained periods from the GLSPs and ACF method to be smaller than 10\%. 
    {
    An initial sample of 184 periodic quasars candidates is obtained. 
    As a second step, we evaluate the significance of the periodicity by calculating the FAP and comparing the peak power with the background red-noise power levels in the GLSP. We adopt the null hypothesis that quasar stochastic variabilities arise from red-noises modeled by either DRW or SPL. We set a limitation of $\rm (1-FAP)\geq99.73\%$ and require the highest GLSP peak exceeding the 3$\sigma$ level. 
    Additionally, for the case of DRW red-noises, we can directly fit the light curves and calculate the BIC to compare between the pure stochastic DRW model and the periodic (sinusoidal + DRW) model. To select out the sinusoidal + DRW model, we adopt a criterion of $\rm \Delta BIC \leq -10$. Finally, we identify 106 (DRW) and 86 (SPL) significant periodic candidates out of a total of 143,700 quasars, respectively. The Bayes factors derived from {\tt RECON} demonstrate that for our initial 184 candidates, the SPL model is relatively preferable over the DRW model (see Section~\ref{sec_competition}). Therefore, we choose the sample of 86 periodic candidates selected using the SPL model as our final candidate sample. 
    
    We perform Monte Carlo simulations to estimate the global look-elsewhere effect, which accounts for false positives resulting from a global search over an AGN sample due to random fluctuations (see Section~\ref{sec_global_fap}). The obtained ``global FAP'' is $\sim 3.6\times10^{-4}$, meaning that there will be about 51 false positives in a sample of 143,700 quasars. A such FAP is comparable to the probability of discovering the above 86 periodic candidates, indicating that, statistically, most of the periodic candidates in our sample might be normal AGNs with their periodic variability caused by red-noise.

    Notably, compared to the previous searches made by \cite{Graham2015b,Charisi2016}, their parent quasar catalogs are significantly overlapped with our catalog, however, we find that the identified periodic quasar candidates are not matching.
    In addition, we extend the ZTF light curves of the 86 candidates by supplying other earlier survey data (including CRTS, PTF, PS1, and ATLAS) to verify the periodicity. With extended light curves, only three quasars with 6-8 cycles of periods satisfy the selection criteria (see Section~\ref{sec_periodicity_verify}). These results, together with the high ``global FAP'' mentioned above, impose an implication that the variations of the candidates are most likely quasi-periodic rather than strictly periodic or that they are purely caused by stochastic red-noise. In this case, the selection criteria and methods used in both this work and previous works  (e.g., \citealt{Graham2015b,Charisi2016,Liu2019,Chen2020}) might be no longer effective. Indeed, those criteria are inclined to identify candidates with sinusoidal variations. In summary, more time-domain data of all identified candidates with longer temporal baselines are necessary to confirm their periodicity and rule out false positives due to red-noise. Also, new effective methods are highly warranted to identify generic quasi-periodic quasar candidates.
    
    } 
	
    \section*{Acknowledgements}
    We thank the referee for instructive comments that significantly improved the manuscript.
    This research is supported by National Key R\&D Program of China (2021YFA1600404); by NSFC-11991050, -11833008, -11991052, -11873048, -12022301, -11991051, -11991054; by the Key Research Program of Frontier Sciences,
    CAS, grant QYZDJ-SSW-SLH007 and by the China Manned Space Project with no. CMS-CSST-2021-A06.
    Y.-R.L. acknowledges financial support from the NSFC grants 11922304 and 12273041, from the Youth Innovation Promotion Association CAS.
    
    Based on observations obtained with the Samuel Oschin Telescope 48-inch and the 60-inch Telescope at the Palomar
    Observatory as part of the Zwicky Transient Facility project. ZTF is supported by the National Science Foundation under Grant
    No. AST-2034437 and a collaboration including Caltech, IPAC, the Weizmann Institute for Science, the Oskar Klein Center at
    Stockholm University, the University of Maryland, Deutsches Elektronen-Synchrotron and Humboldt University, the TANGO
    Consortium of Taiwan, the University of Wisconsin at Milwaukee, Trinity College Dublin, Lawrence Livermore National
    Laboratories, and IN2P3, France. Operations are conducted by COO, IPAC, and UW.

    {The CSS survey is funded by the National Aeronautics and Space Administration under grant No. NNG05GF22G issued through the Science Mission Directorate Near-Earth Objects Observations Program. The CRTS survey is supported by the U.S. National Science Foundation under grants AST-0909182.

    The Pan-STARRS1 Surveys (PS1) and the PS1 public science archive have been made possible through contributions by the Institute for Astronomy, the University of Hawaii, the Pan-STARRS Project Office, the Max-Planck Society and its participating institutes, the Max Planck Institute for Astronomy, Heidelberg and the Max Planck Institute for Extraterrestrial Physics, Garching, The Johns Hopkins University, Durham University, the University of Edinburgh, the Queen's University Belfast, the Harvard-Smithsonian Center for Astrophysics, the Las Cumbres Observatory Global Telescope Network Incorporated, the National Central University of Taiwan, the Space Telescope Science Institute, the National Aeronautics and Space Administration under Grant No. NNX08AR22G issued through the Planetary Science Division of the NASA Science Mission Directorate, the National Science Foundation Grant No. AST-1238877, the University of Maryland, Eotvos Lorand University (ELTE), the Los Alamos National Laboratory, and the Gordon and Betty Moore Foundation.

    This paper is based on observations obtained with the Samuel
    Oschin Telescope as part of the PTF project, a scientific collaboration between the California Institute of Technology, Columbia
    University, Las Cumbres Observatory, the Lawrence Berkeley
    National Laboratory, the National Energy Research Scientific Computing Center, the University of Oxford, and the Weizmann Institute
    of Science.

    This work has made use of data from the Asteroid Terrestrial-impact Last Alert System (ATLAS) project. The Asteroid Terrestrial-impact Last Alert System (ATLAS) project is primarily funded to search for near earth asteroids through NASA grants NN12AR55G, 80NSSC18K0284, and 80NSSC18K1575; byproducts of the NEO search include images and catalogs from the survey area. This work was partially funded by Kepler/K2 grant J1944/80NSSC19K0112 and HST GO-15889, and STFC grants ST/T000198/1 and ST/S006109/1. The ATLAS science products have been made possible through the contributions of the University of Hawaii Institute for Astronomy, the Queen’s University Belfast, the Space Telescope Science Institute, the South African Astronomical Observatory, and The Millennium Institute of Astrophysics (MAS), Chile.
    
    Funding for the Sloan Digital Sky Survey IV has been provided by the Alfred P. Sloan Foundation, the U.S. Department of Energy Office of Science, and the Participating Institutions. SDSS-IV acknowledges support and resources from the Center for High-Performance Computing at the University of Utah.
    The SDSS web site is \url{www.sdss.org}. 
    SDSS is managed by the Astrophysical Research Consortium for the Participating Institutions of the SDSS Collaboration, including the Carnegie Institution for Science, Chilean National Time Allocation Committee (CNTAC) ratified researchers, the Gotham Participation Group, Harvard University, Heidelberg University, The Johns Hopkins University, L’Ecole polytechnique f\'{e}d\'{e}rale de Lausanne (EPFL), Leibniz-Institut f\"{u}r Astrophysik Potsdam (AIP), Max-Planck-Institut f\"{u}r Astronomie (MPIA Heidelberg), Max-Planck-Institut f\"{u}r Extraterrestrische Physik (MPE), Nanjing University, National Astronomical Observatories of China (NAOC), New Mexico State University, The Ohio State University, Pennsylvania State University, Smithsonian Astrophysical Observatory, Space Telescope Science Institute (STScI), the Stellar Astrophysics Participation Group, Universidad Nacional Aut\'{o}noma de M\'{e}xico, University of Arizona, University of Colorado Boulder, University of Illinois at Urbana-Champaign, University of Toronto, University of Utah, University of Virginia, Yale University, and Yunnan University.
	
	\section*{Data Availability}
	The data underlying this article will be shared on reasonable request to the corresponding author.

	
	
	\bibliographystyle{mnras}
	\bibliography{reference} 

	
	
	
	\appendix    
     \section{The Effect of Irregular Sampling on PSDs}
     \label{sec_psd_test}

     \begin{figure}
    \centering
    \includegraphics[width=0.43\textwidth]{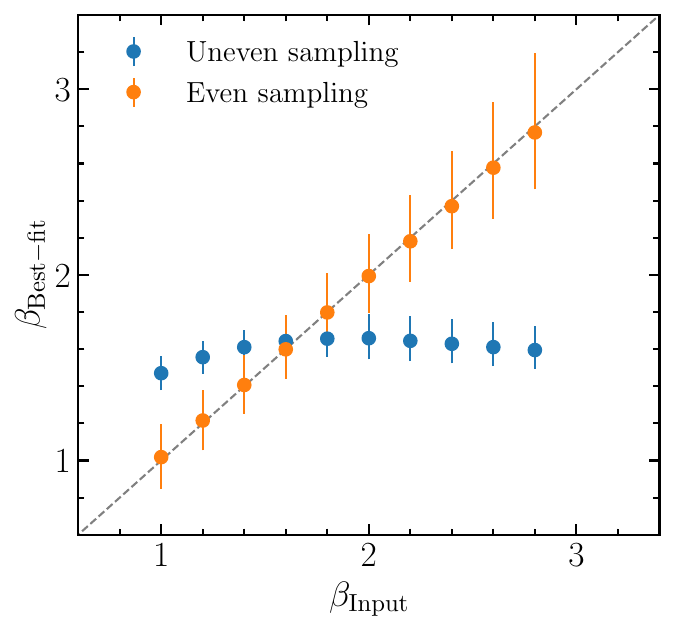}
    \caption{Recovered SPL slopes by directly fitting PSDs from evenly/unevenly sampled light curves versus the input slopes.}
    \label{fig_recovery_beta}
    \end{figure}
    
     \begin{figure*}
    \centering
    \includegraphics[width=0.95\textwidth]{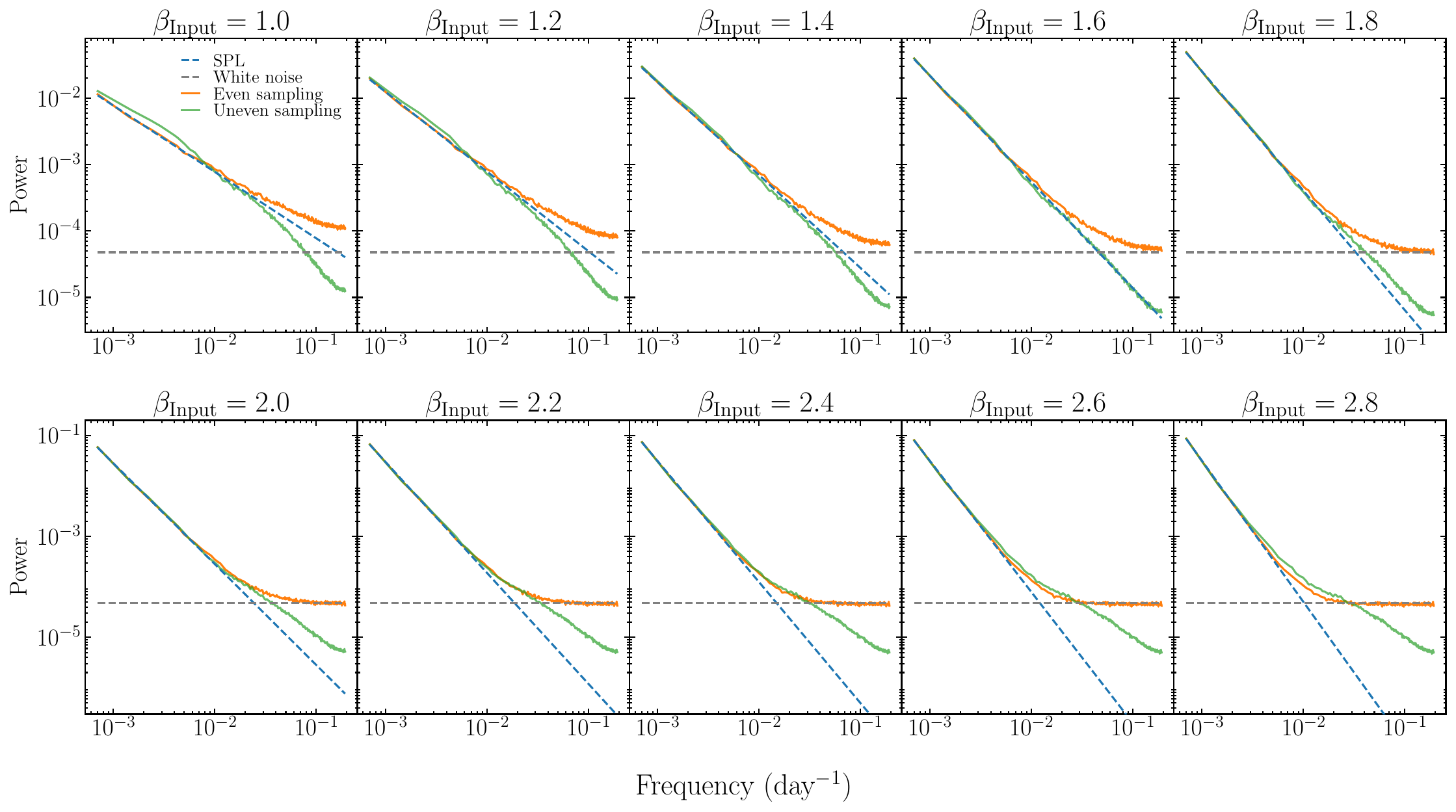}
    \caption{Comparison of calculated PSDs from even and uneven mock light curves. The horizontal dashed lines represent the white noise power from injected errors. }
    \label{fig_psd_comparison}
    \end{figure*}    

     \begin{figure}
    \centering
    \includegraphics[width=0.48\textwidth]{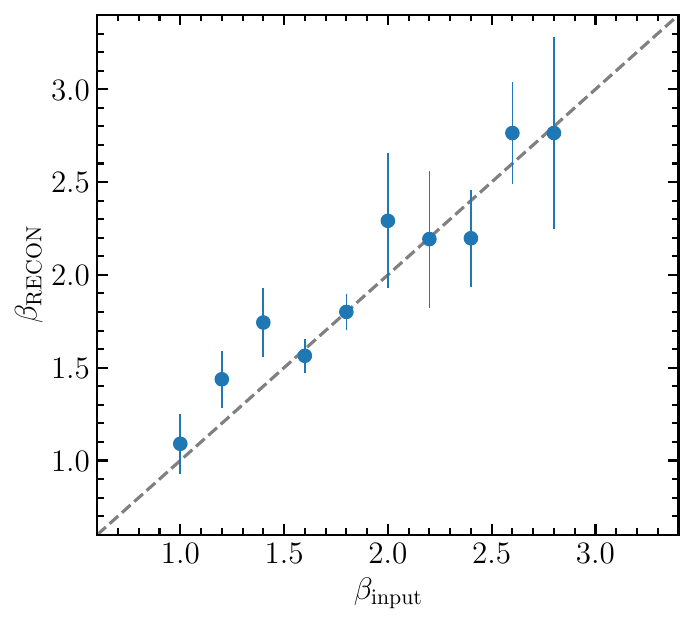}
    \caption{Recovered SPL slopes using {\tt RECON} versus the input values.}
    \label{fig_beta_recon}
    \end{figure}
    
    { 
     To calculate PSDs of irregularly sampled light curves, we usually need to resample the light curves into an even time grid through interpolation. However, such an interpolation manipulation may significantly distort calculated PSDs from the true spectra. To illustrate this point, we conduct a simulation test as follows.
     
     First, we generate mock light curves with daily sampling using the method of \cite{Timmer1995}. We adopt a SPL PSD and set the slope $\beta$ in a range of (1.0-2.8) with a step of $\Delta\beta=0.2$. We down-sample the mock light curves to be consistent with the sampling of the observed light curve of a randomly selected quasar. This yields an uneven light curve.
     To generate an even light curve, we set an even time grid the same as that used to calculate the PSD of the observed light curve with resampling. We then linearly interpolate the mock daily light curve onto this time grid. Gaussian noises are added to both the evenly and unevenly sampled light curves, with a zero mean and standard deviation equal to the mean photometric uncertainty of the observed light curve. Finally, we calculate the PSDs and directly fit them to determine the corresponding slopes.
     
     We repeat the above procedures 1,000 times. Fig.~\ref{fig_recovery_beta} shows a comparison between the recovered and input slopes. For the case of evenly sampled light curves, the recovered slopes are well-consistent with the inputs within errors. While for the uneven cases, the recovered slopes significantly deviate from the inputs, especially when $\beta>2$. 
     This is because the interpolation of uneven light curves leads to significant distortions to the PSDs at high frequencies. Furthermore, for each $\beta$, we calculate the mean PSDs from the simulated even and uneven  light curves in Fig.~\ref{fig_psd_comparison}. As expected, there appear significant differences at higher frequencies ($f>10^{-2}\ \rm day^{-1}$).

     Alternatively, we employ the {\tt RECON} framework (\citealt{Li2018}; also see a brief introduction in Section~\ref{sec_fap_pl}) to determine the SPL parameters for the above generated mock light curves. Considering the heavy time consumption of running {\tt RECON}, we randomly select ten mock light curves with different slopes. The obtained slopes shown in Fig.~\ref{fig_beta_recon} are generally consistent with the input values within 2$\sigma$ uncertainties.  This test demonstrates that the {\tt RECON} framework can better infer the PSD parameters from unevenly sampled light curves.}

	\section{Spectroscopy and Quasar properties} \label{sec_spec}

    \begin{figure}
        \centering
        \includegraphics[width=0.48\textwidth]{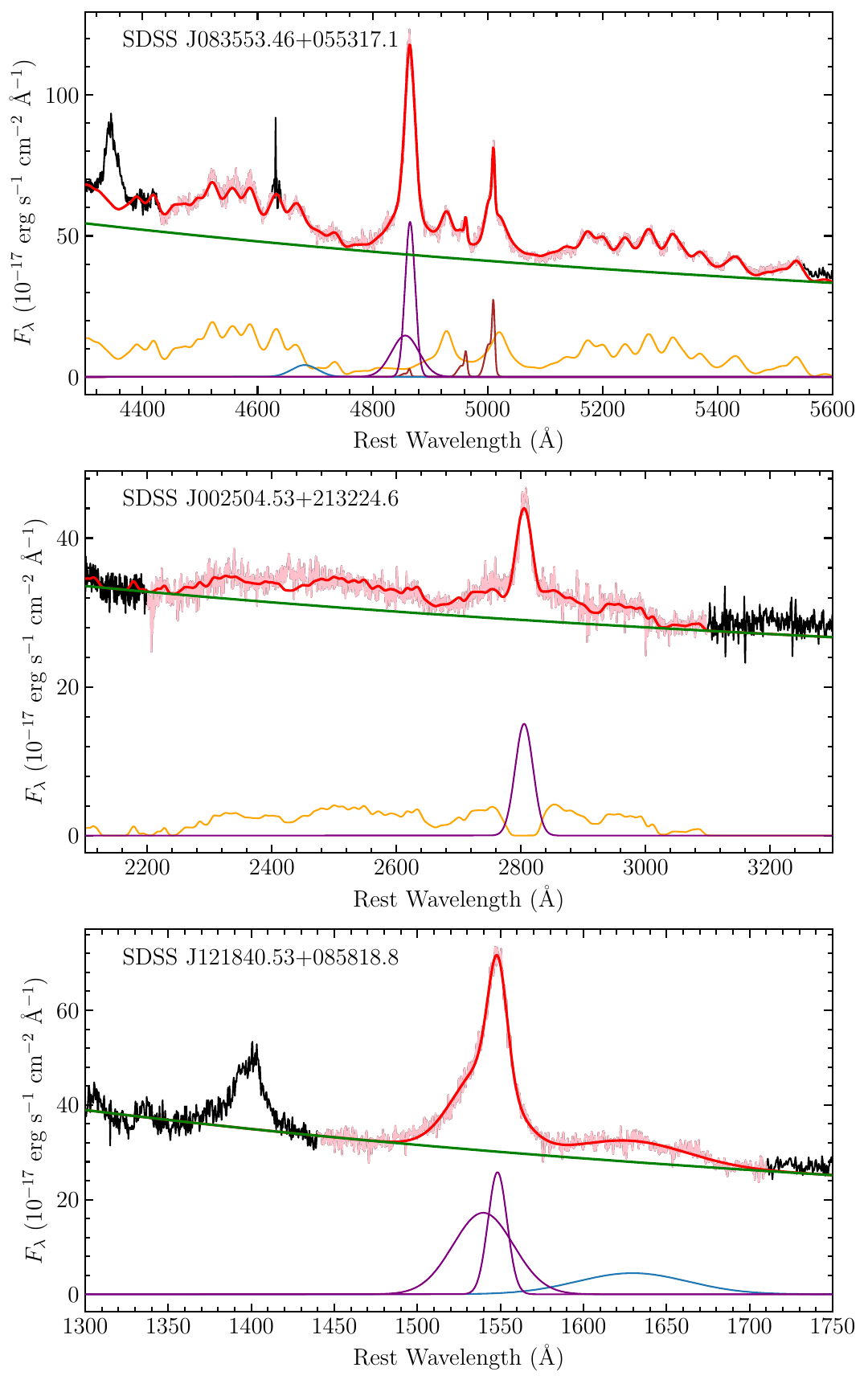}
        \caption{Examples of spectral decomposition for the H$\beta$, {\mgii}, and {\civ} lines (from top to bottom), respectively. The black color shows the spectral data and the pink color shows the fitting window. The red solid lines show the best fit.}
        \label{fig_fit}
    \end{figure}
	We measure the AGN properties of our candidates using spectroscopic data from the SDSS Science Archive Server. We retrieved archival spectra for 178 candidates and discard the spectrum of SDSS J225830.24+213118.2 due to its poor data quality.

	To estimate the black hole mass, we need the widths of the broad emission line and continuum luminosity. As usual, we adopt the broad $\rm H\beta$, {\mgii} or {\civ} lines according to the redshift of the candidate.
	For the H$\beta$ line, we fit the spectra in a wavelength window 4400-5500 {\AA} using the following components: (1) a simple power law for continuum, (2) one or two Gaussians with the same velocity width and shift for the narrow lines including H$\beta$ and {\oiii}$\lambda\lambda4959, 5007$, (3) a Gaussian for the narrow and broad \heii$\ \lambda4686$ lines if needed, (4) one or two Gaussians for the broad H$\beta$ line, (5) the Fe II template from \cite{Boroson1992}, (6) the stellar template from \cite{Bruzual2003} with an age of 11 Gyr and a metallicity of $Z$ = 0.05 to account for the host galaxy starlight if the spectra show apparent absorption features. For those spectra with very weak {\oiii}$\lambda\lambda4959, 5007$ doublet, the narrow H$\beta$ line will not be included in the above spectral decomposition considering that the narrow H$\beta$/\oiii$\ \lambda5007$ flux ratio is typically about 0.1 in AGNs \citep{Veilleux1987}. We measure the FWHM of the broad $\rm H\beta$ line from the fitted two Gaussians and the flux density $F_\lambda$ at 5100 {\AA} from  the power-law continuum. We estimate the black hole mass using the equation
	\begin{equation}\label{eq5}
		M_{\bullet}=f\frac{{\rm V}^2_{\rm H\beta}R_{\rm H\beta}}{G},
	\end{equation}
	where $f$ is the virial factor, $G$ is the gravitational constant, $V_{\rm H\beta}$ is the FWHM of the broad H$\beta$ line and $R_{\rm H\beta}$ is the size of the  H$\beta$ BLR. We adopt $f=1.12\pm0.31$ from \cite{Woo2015}. $R_{\rm H\beta}$ is calculated using the size-luminosity relationship (\citealt{Bentz2013})
	\begin{equation}\label{eq6}
		\log\left(\frac{R_{\rm H\beta}}{\rm 1 lt-day} \right) = K + \alpha \log\left(\frac{L_{5100}}{\rm 10^{44}\ erg\ s^{-1}}\right),
	\end{equation}
	where the coefficients $K=1.527^{+0.031}_{-0.031}$ and $\alpha=0.533^{+0.035}_{-0.033}$, and $L_{5100}$ is the luminosity at 5100~{\AA}. With Equation~(\ref{eq6}), Equation~(\ref{eq5}) is recast into
	\begin{equation}
		\begin{aligned}
			\log\left(\frac{M_{\bullet}}{M_{\odot}}\right)=&\log \left[ \left(\frac{\rm FWHM(H\beta)}{\rm\ 5000\ km\ s^{-1}}\right)^2 \left(\frac{L_{5100}}{\rm 10^{44}\ erg\ s^{-1}}\right)^{0.53} \right]\\& + 8.27.
		\end{aligned}
	\end{equation}
	
	For the {\mgii} line, we choose a wavelength window 2200-3100~{\AA}. We include: 1)a single power law for the AGN continuum, 2) a single Gaussian for the {\mgii} line and the Fe II template from \cite{Vestergaard2001}. For those spectra with low S/N, the Fe II template is not included to avoid the overfitting and subtracting \citep{Liu2019}. Meanwhile, the narrow {\mgii} line is also not included because it is highly degenerated with the {\mgii}$\lambda\lambda2795.5, 2802.7$ doublet, so it is difficulty to decompose reliably. We extract the FWHM of {\mgii} from the best-fit Gaussian and determine the continuum luminosity at 3000 {\AA} ($L_{3000}$) from the power-law component. The black hole mass is estimated using the equation from \citet{Shen2011},
	\begin{equation}
		\begin{aligned}
			\log\left(\frac{M_{\bullet}}{M_{\odot}}\right) =&\log\left[\left(\frac{\rm FWHM(\mbox{\rm \mgii})}{\rm 5000\ km\ s^{-1}}\right)^2 \left(\frac{L_{3000}}{\rm 10^{44}\ erg\ s^{-1}}\right)^{0.62}\right]\\&+8.14.
		\end{aligned}
	\end{equation}
	
	Finally, for the {\civ} line, we use a wavelength window 1440-1710~{\AA} and mask the absorption lines. We include a single power law for the continuum, two Gaussians for {\civ} and a single Gaussian for $\rm\heii\ \lambda1640$. Again, the narrow {\civ} line is not included due to its degeneracy with the {\civ}$\lambda\lambda1548, 1551$ doublet. We measured the  {\civ} FWHM from the two best-fit Gaussian and the luminosity at 1450 {\AA} ($L_{1450}$) from the power-law continuum. We estimate the black hole mass using the equation from \citet{Vestergaard2006},
	\begin{equation}
		\begin{aligned}
			\log\left(\frac{M_{\bullet}}{M_{\odot}} \right) =& \log \left[ \left(\frac{\rm FWHM(\mbox{\rm \civ})}{\rm 6000\ km\ s^{-1}}\right)^2 \left(\frac{L_{1450}}{\rm 10^{44}\ erg\ s^{-1}}\right)^{0.53} \right] \\& + 8.22.
		\end{aligned}
	\end{equation}

	With the black hole mass, we estimate the dimensionless accretion rate as $\dot{\mathscr{M}}$ \citep{Du2014}
	\begin{equation}
		\dot{\mathscr{M}} = \frac{\dot{M}}{L_{\rm Edd}c^{-2}} = 20.1 \left(\frac{\ell_{44}}{\cos i}\right)^{3/2} m^{-2}_7,
	\end{equation}
	where $\dot M$ is the mass accretion rate, $L_{\rm Edd}$ is the Eddington luminosity, $\ell_{44} = L_{5100} / 10^{44}\ {\rm erg\ s^{-1}}$, $m_7 = M_{\bullet} / 10^7 M_{\odot}$ and $i$ is the inclination of disk (we adopt $\cos i\approx0.75$). We convert the monochromatic luminosity from 1450~{\AA} or 3000~{\AA} to 5100~{\AA} with the bolometric correction factors from \citet{Runnoe2012}.

	Fig.~\ref{fig_fit} shows three examples of spectral decomposition for the H$\beta$, {\mgii}, and {\civ} lines, respectively. We list $M_{\bullet}$, $L_{5100}$ and dimensionless accretion rates $\dot{\mathscr{M}}$ in Table~\ref{table1}.
	It is worth mentioning two points regarding the black hole mass estimation. First, the black hole mass obtained using {\civ} and {\mgii} estimator have a lager systematic uncertainty of $\sim$0.3 dex (\citealt{Vestergaard2006, Shen2011}) and the $R_{\rm H\beta}-L_{5100}$ relation for H$\beta$ estimator has an intrinsic scatter of $\sim$0.19 dex \citep{Bentz2013}. We include these systematic uncertainties when calculating the black hole mass. Second, however, when converting the 1450~{\AA} and 3000~{\AA} luminosity to 5100~{\AA} luminosity, we do not include uncertainties of the bolometric correction factors, which are indeed not well determined.

	{
	\section{The Extended Light Curves of 86 selected periodic candidates}\label{sec_lc}

    In Fig.~\ref{fig_lc_extend_all}, we plot the extended light curves of the 86 periodic candidates.
    }


	\bsp	
	\label{lastpage}

    \begin{figure*}
        \centering
        \includegraphics[width=0.94\textwidth]{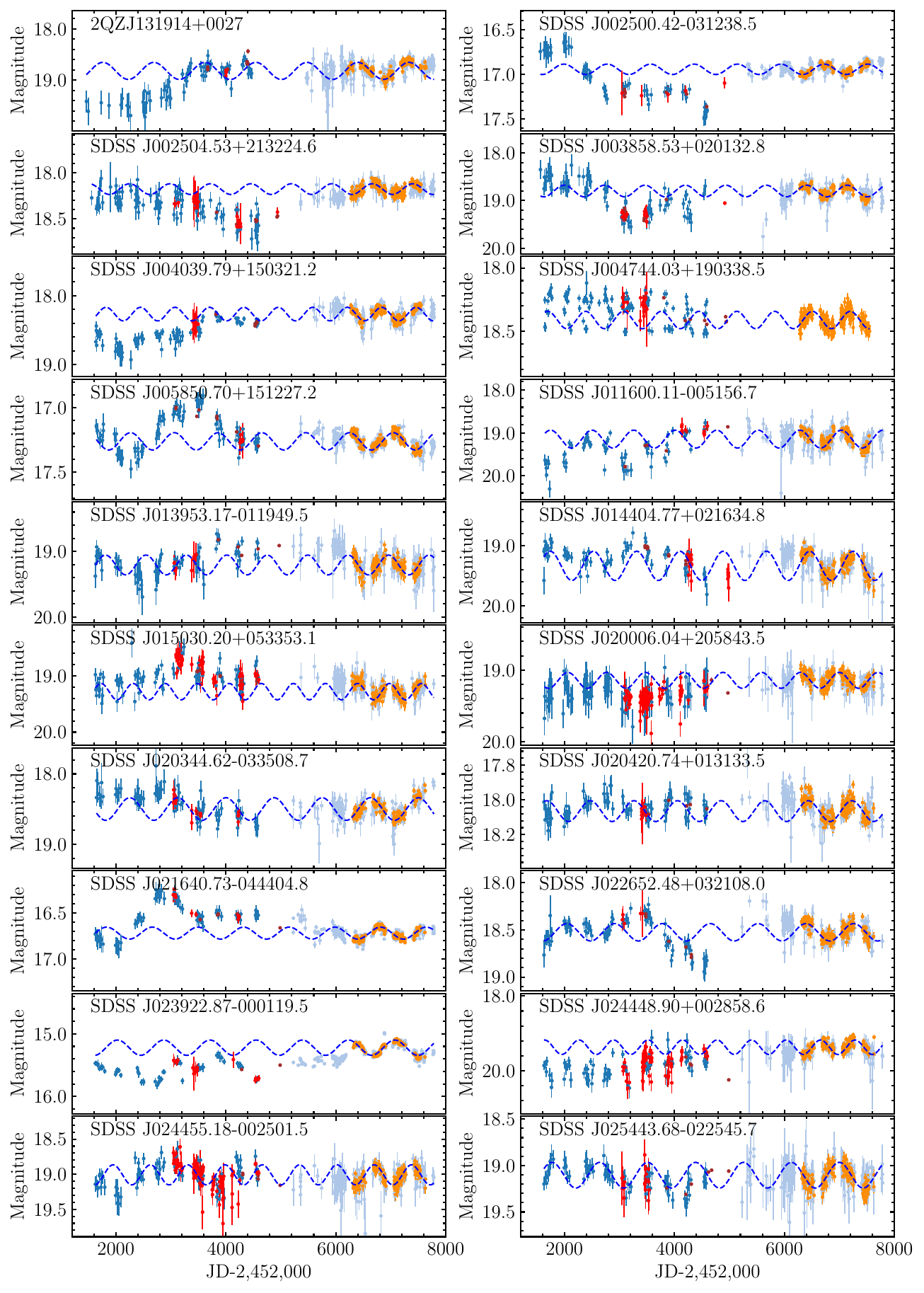}
        \caption{The synthetic light curves of 86 candidates from the CRTS (blue), PS1 (brown), PTF (red), ATLAS (light blue), and ZTF (orange) archival survey data. The blue dash lines show the best sinusoidal fits for the binned ZTF light curves (see Section~\ref{sec_method} for a detail).}
        \label{fig_lc_extend_all}
    \end{figure*}
    \begin{figure*}
        \centering
        \includegraphics[width=0.94\textwidth]{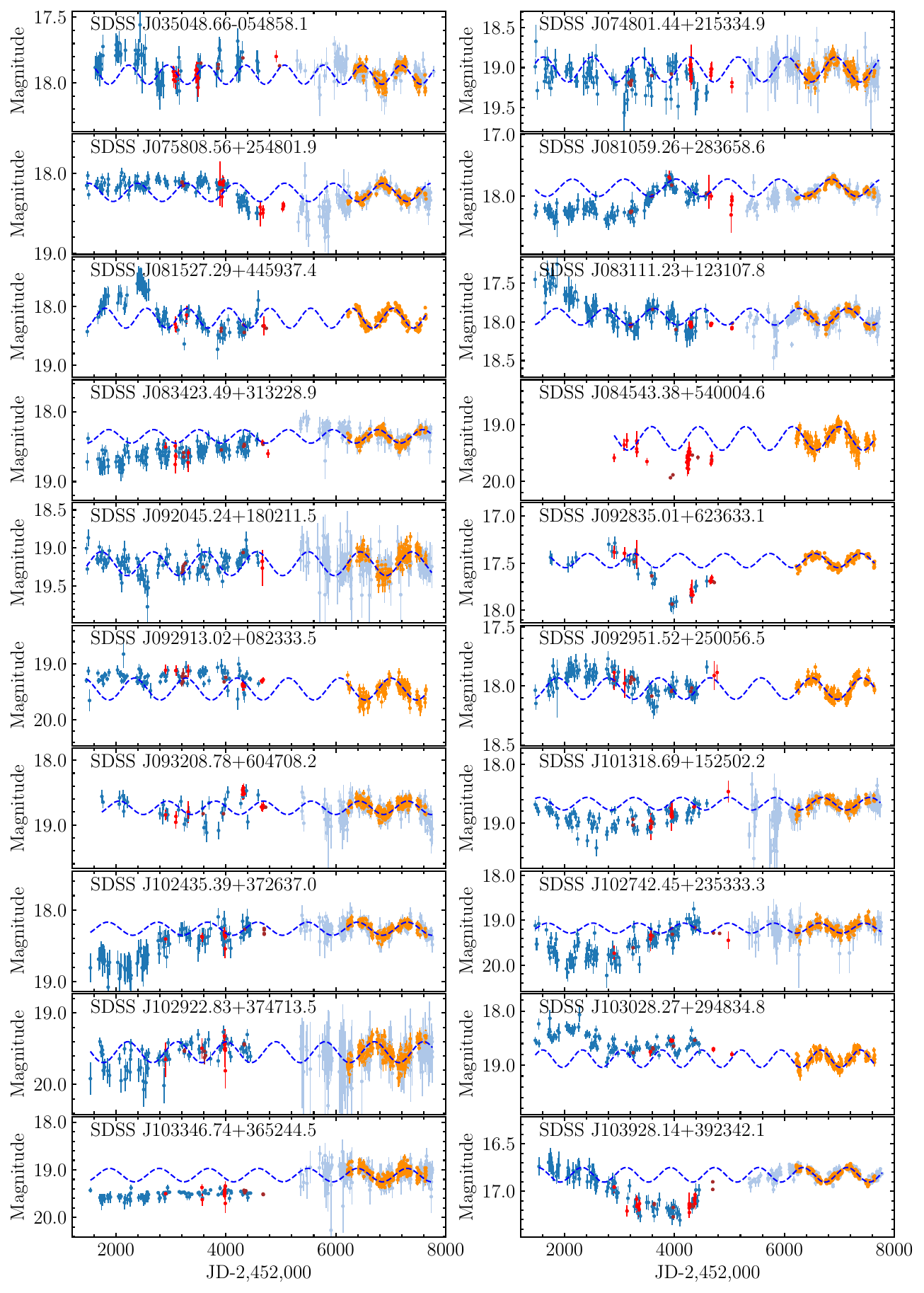}
        \addtocounter{figure}{-1}
        \caption{ (Continued).}
        \label{fig_lc_extend_all}
    \end{figure*}
    \begin{figure*}
        \centering
        \includegraphics[width=0.94\textwidth]{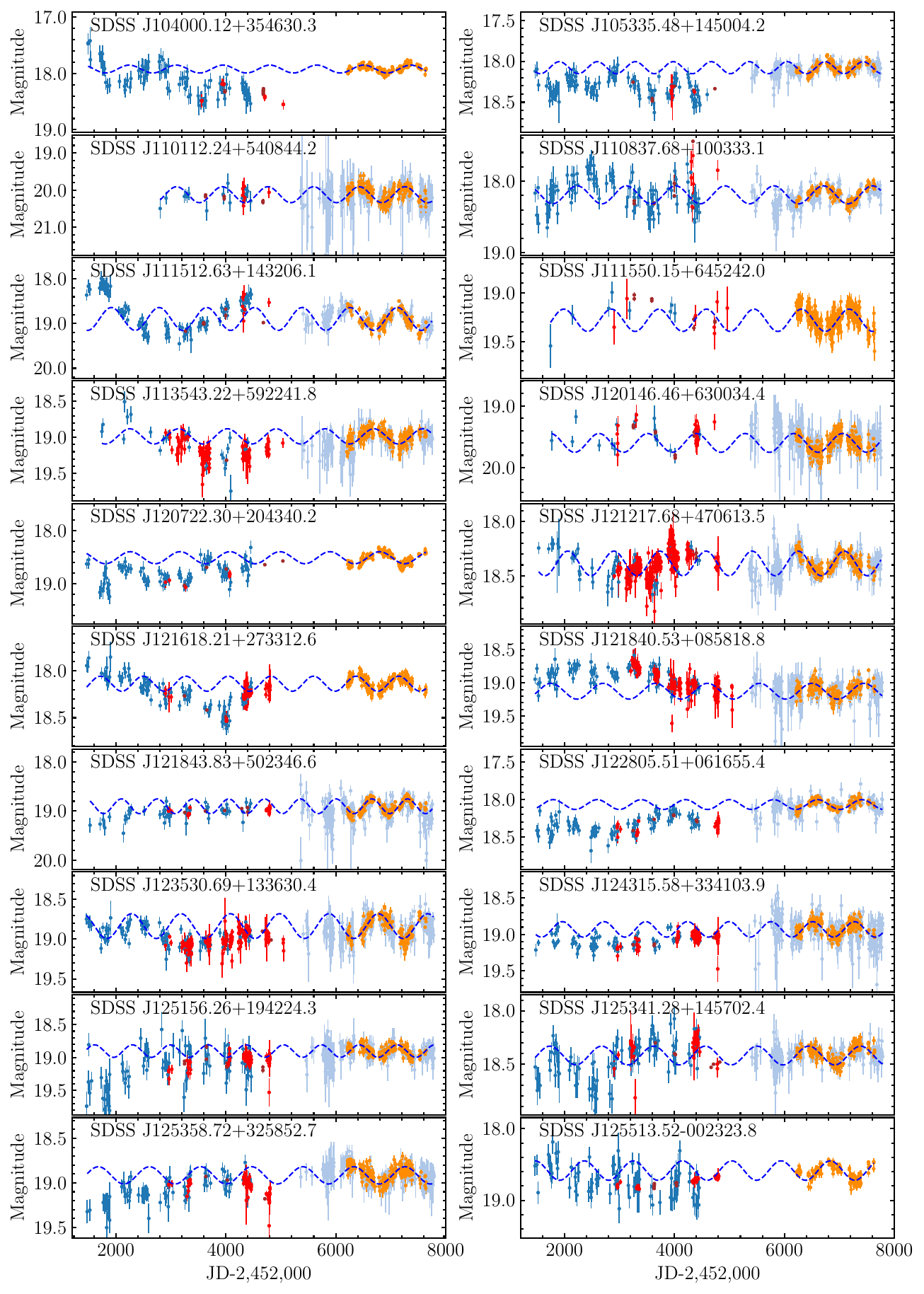}
        \addtocounter{figure}{-1}
        \caption{(Continued).}
        \label{fig_lc_extend_all}
    \end{figure*}
    \begin{figure*}
        \centering
        \includegraphics[width=0.94\textwidth]{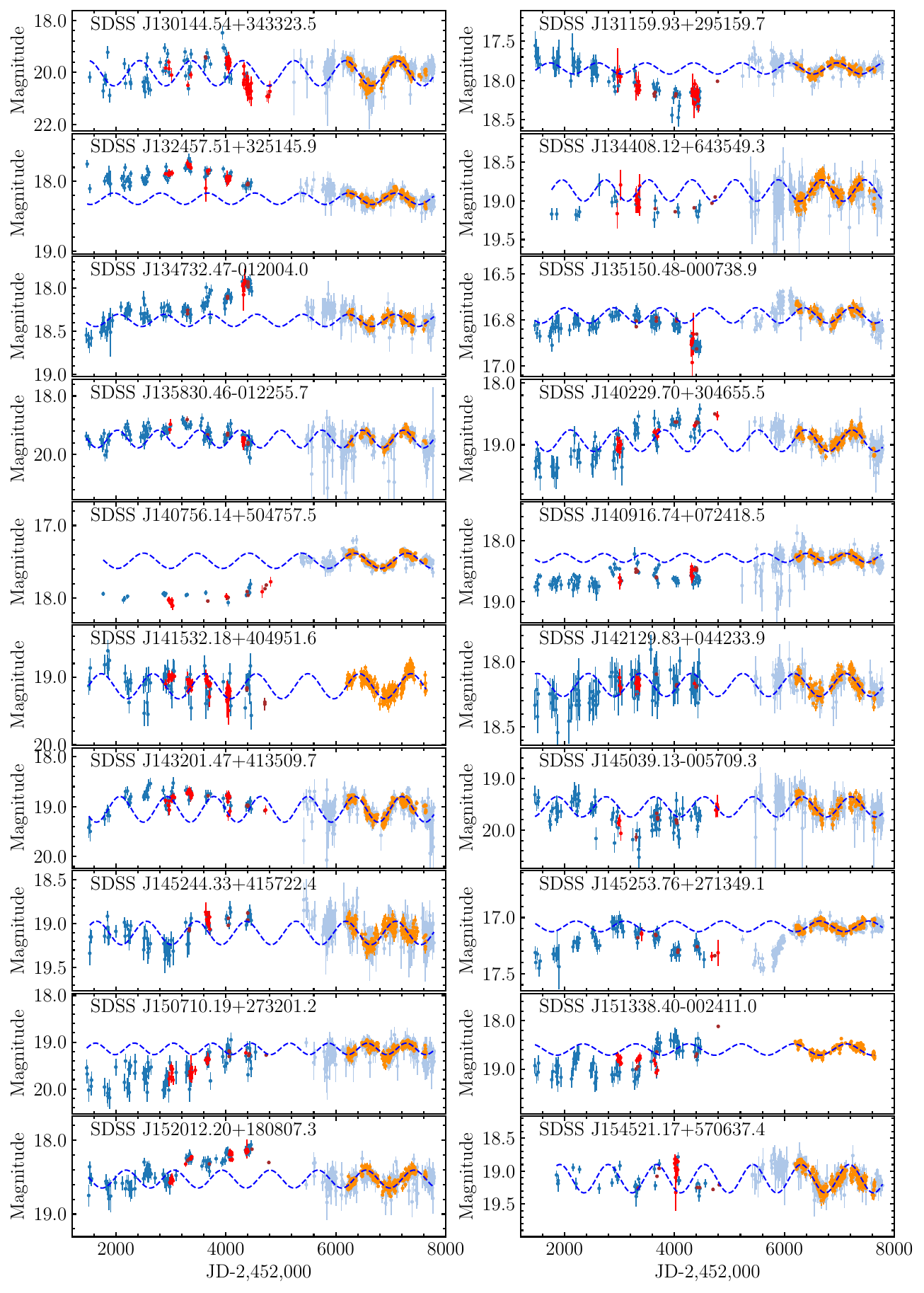}
        \addtocounter{figure}{-1}
        \caption{(Continued).}
        \label{fig_lc_extend_all}
    \end{figure*}
    \begin{figure*}
        \centering
        \includegraphics[width=0.94\textwidth]{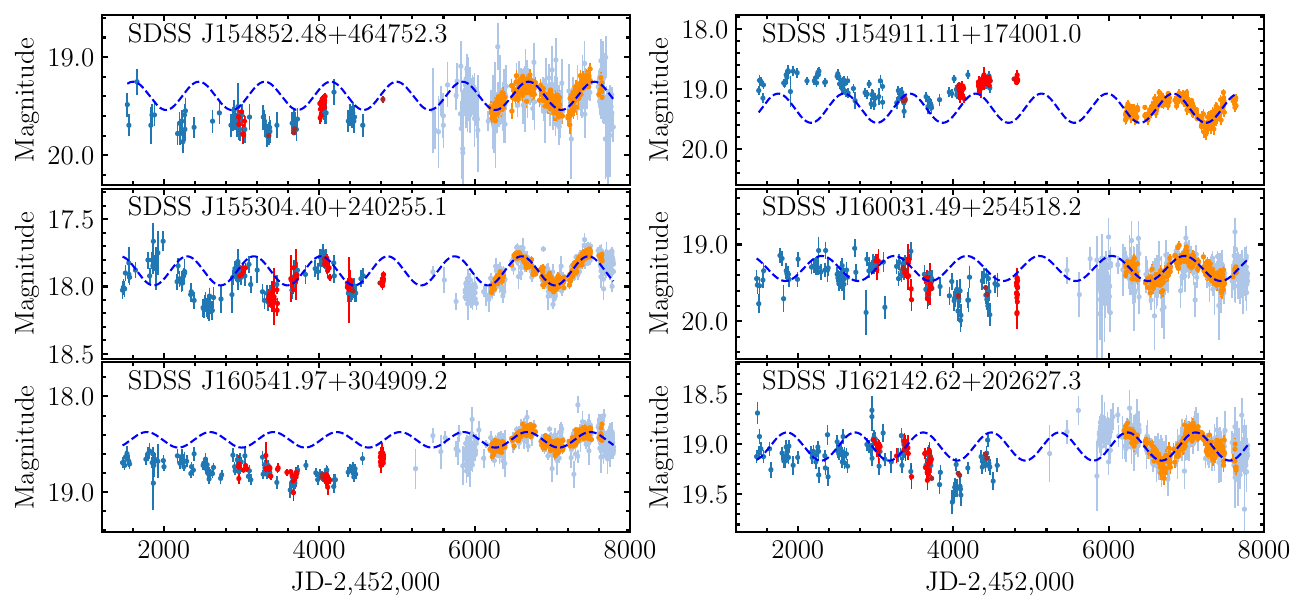}
        \addtocounter{figure}{-1}
        \caption{(Continued).}
        \label{fig_lc_extend_all}
    \end{figure*}

\end{document}